\documentclass[man,floatsintext]{apa7}

\usepackage[american]{babel}
\usepackage{amsmath}
\usepackage{amssymb}
\usepackage{bm}
\usepackage{booktabs}
\usepackage{graphicx}
\usepackage{array}
\usepackage{makecell}
\usepackage{tabularx}
\usepackage{threeparttable}
\usepackage{xcolor}
\usepackage{listings}
\usepackage{upquote}
\usepackage{xurl}
\usepackage{csquotes}
\usepackage[section]{placeins}
\usepackage[style=apa,sortcites=true,sorting=nyt,backend=biber]{biblatex}
\DeclareLanguageMapping{american}{american-apa}
\hypersetup{hidelinks}
\counterwithout{equation}{subsection}

\definecolor{codeframe}{RGB}{180,180,180}

\providecommand{\nolinkurl}[1]{\url{#1}}
\DeclareRobustCommand{\Rfunc}[1]{\nolinkurl{#1}}
\DeclareRobustCommand{\Rarg}[1]{\nolinkurl{#1}}
\DeclareRobustCommand{\Robj}[1]{\nolinkurl{#1}}
\DeclareRobustCommand{\Rpkg}[1]{\nolinkurl{#1}}
\newcommand{\apafiguregraphic}[1]{\includegraphics[width=\linewidth,height=.78\textheight,keepaspectratio]{#1}}

\lstdefinestyle{RStyle}{
  language=R,
  basicstyle=\ttfamily\footnotesize,
  frame=single,
  rulecolor=\color{codeframe},
  framerule=0.4pt,
  xleftmargin=0.5em,
  xrightmargin=0.5em,
  aboveskip=0.6em,
  belowskip=0.6em,
  keepspaces=true,
  showstringspaces=false,
  columns=fullflexible,
  breaklines=true,
  breakatwhitespace=false,
  tabsize=2,
  upquote=true,
  commentstyle=\itshape,
  keywordstyle=\bfseries,
}

\lstdefinestyle{ROutStyle}{
  language={},
  basicstyle=\ttfamily\footnotesize,
  frame=single,
  rulecolor=\color{codeframe},
  framerule=0.4pt,
  keepspaces=true,
  columns=fullflexible,
  showstringspaces=false,
  tabsize=2,
  xleftmargin=0.5em,
  xrightmargin=0.5em,
  breaklines=true,
  breakatwhitespace=false,
  aboveskip=0.6em,
  belowskip=0.6em,
  upquote=true,
}

\title{GGMNIRA as a Gaussian Graphical Model Extension of the NodeIdentifyR Algorithm for Projected Node Importance}
\shorttitle{GGMNIRA FOR PROJECTED NODE IMPORTANCE}

\authorsnames[1,{2,3*},{1*}]{Yiming Wu, Fei Wang, Hongyun Liu}

\authorsaffiliations{
  {Faculty of Psychology, Beijing Normal University, Beijing, 100875, China},
  {State Key Laboratory of Cognitive Science and Mental Health, Institute of Psychology, Chinese Academy of Sciences, 100101, Beijing, China},
  {Department of Psychology, University of Chinese Academy of Sciences, 100049, Beijing, China}
}

\leftheader{Wu, Wang, and Liu}

\authornote{
\textsuperscript{*}Fei Wang and Hongyun Liu are co-corresponding authors.

Correspondence concerning this article should be addressed to Fei Wang,
State Key Laboratory of Cognitive Science and Mental Health, Institute of Psychology,
Chinese Academy of Sciences, No. 16 Lincui Road, Chaoyang District, Beijing, China.
E-mail: bjnwangfei0501@outlook.com; or Hongyun Liu,
Faculty of Psychology, Beijing Normal University, Beijing, 100875, China.
E-mail: hyliu@bnu.edu.cn.

All data, analysis code, and reproducible materials for this article are available at
\url{https://github.com/ywu6888}.

The authors received no financial support for the research, authorship, or publication of this article.
The authors declare no competing interests.
}

\abstract{

\noindent\textbf{Scientific Abstract:}
The NodeIdentifyR Algorithm (NIRA) has been increasingly applied in psychological network research as a simulated-manipulation approach for comparing the network-level impact associated with different nodes. However, NIRA is restricted to binary variables, and the theoretical interpretation of the manipulated node intercept is primarily grounded in the symptom-activation context of psychopathology. To address these limitations, we extend NIRA to Gaussian graphical model and propose the Gaussian Graphical Model NodeIdentifyR Algorithm (GGMNIRA). GGMNIRA systematically manipulates the conditional mean of each node and uses Kullback--Leibler (KL) divergence to quantify the resulting change in the joint network distribution.  Because the conditional mean has a meaningful interpretation for variables across different areas of psychology, GGMNIRA can be applied to a broader range of psychological research contexts. We also developed complementary procedures for evaluating the uncertainty of GGMNIRA results, including a correlation-stability coefficient and a nonparametric bootstrap difference test for KL divergence. In addition, GGMNIRA was extended to bridge Gaussian graphical model, allowing simulated manipulation to be conducted at both the node and construct levels in multi-construct networks. All methods are implemented in the R package \Rpkg{GGMNIRA}.

\par\medskip

\noindent\textbf{Translational Abstract:}
Researchers using psychological networks often want to know which variables, such as symptoms, personality traits, emotions, or beliefs, may have greater consequences for the network if they are changed. NodeIdentifyR Algorithm (NIRA) provides a projected answer to this question from the perspective of simulated manipulation. However, NIRA is limited to binary variables and was originally developed mainly for symptom networks. In this study, we extend NIRA to Gaussian graphical model and develop Gaussian Graphical Model NodeIdentifyR Algorithm (GGMNIRA), allowing this simulated-manipulation approach to be applied more broadly to psychological variables across different research areas. GGMNIRA estimates how much the overall network is projected to change when the expected level of a given variable is systematically altered. We also provide procedures for assessing the uncertainty of these results and extend the framework to networks containing multiple psychological constructs, where researchers can examine manipulations at either the node or construct level. These results should be interpreted as model-based expectations rather than confirmed causal effects. The methods are implemented in the R package \Rpkg{GGMNIRA}.
}


\begin{document}
\maketitle

Over the past two decades, network analysis has gradually become one of the most active methodological paradigms in psychological research. Unlike traditional approaches that treat psychological variables as passive manifestations of a common latent variable, network analysis directly models psychological variables as an interacting system. In this framework, each variable is represented as a node\footnote{In what follows, we often use the terms ``variable'' and ``node'' interchangeably.}, and the edges between nodes represent unique statistical associations between two variables after controlling for all other variables \parencite{borsboom2021network,borsboom2013network}. This perspective has produced substantive applications across multiple subfields of psychology. In psychopathology, individual symptoms of mental disorders such as depression and anxiety, including insomnia, fatigue, and sadness, are organized into symptom networks, in which symptoms activate and maintain one another and jointly constitute the disorder state \parencite{borsboom2017network,robinaugh2020network}. In personality psychology, personality traits are understood as emergent wholes arising from interrelated specific behavioral tendencies, such as ``liking to attend parties'' and ``liking to make friends,'' rather than as passive reflections of an underlying dimension \parencite{cramer2012dimensions}. In social psychology, individuals' specific beliefs or evaluative components regarding a given issue can likewise be modeled as a network, with edges reflecting the extent to which these components mutually reinforce or constrain one another \parencite{dalege2016toward}. In research on cognitive abilities, working memory, processing speed, and other cognitive abilities can be represented as interconnected nodes, which helps explain why different subtests of intelligence tests are generally positively correlated \parencite{van2006dynamical}. The broad application of network analysis in these areas has also promoted the continued methodological development and maturation of this framework. Systematic technical advances have been made in areas ranging from the estimation of network structures and the assessment of parameter stability to the detection of node communities and the comparison of networks across groups \parencite{borsboom2021network,burger2023reporting,epskamp2018tutorial}.

Across these research areas, one core question has consistently attracted researchers' attention: which nodes in the network are most important? In psychopathology, this question directly concerns the prioritization of treatment targets. Researchers aim to identify key symptoms that, once intervened upon, may trigger the broadest cascade of improvement. In personality and social psychology, the same question concerns which traits or beliefs, when changed, may lead to systematic changes in the overall psychological state. The most widely used solution to this problem in psychological network research is to compute \textit{centrality indices} \parencite{cramer2010comorbidity}.

However, as psychological network analysis has evolved, an increasing number of scholars have raised concerns about the applicability of centrality indices in psychological networks. Conceptually, these indices originate from social network analysis, and their underlying assumptions may lack sufficient theoretical justification in psychological contexts \parencite{borgatti2005centrality,bringmann2019centrality}. Statistically, some commonly used indices, particularly \textit{closeness centrality} and \textit{betweenness centrality}, are susceptible to spurious covariance and often show low stability across samples \parencite{epskamp2018estimating,hallquist2021problems}. More fundamentally, centrality indices evaluate node importance primarily on the basis of the topological characteristics of nodes within a network. Yet, according to the definition of node importance proposed by \textcite{dablander2019node}, such an evaluation should instead be grounded in a manipulation-based perspective; that is, the importance of a node should be assessed by manipulating that node and examining the extent to which this manipulation affects the functioning of the network. From this perspective, the most direct approach would be to manipulate the target node in a real-world setting and observe the resulting changes in the network. In practice, isolating the manipulation of a single node is often difficult because nodes may be coupled with one another and cannot be fully separated. Consequently, what is actually manipulated may be a group of nodes rather than a single isolated node.

Against this background, researchers in psychopathology have begun to simulate real-world node manipulations by modifying relevant parameters within statistical models. Specifically, this approach applies computational perturbations to nodes within the model and examines the projected changes in the network resulting from these perturbations. This line of research is commonly referred to as \textit{in silico intervention}.\footnote{Because these methods simulate real-world node manipulations through parameter perturbations within an estimated model, following related formulations by \textcite{van2024measurement} and \textcite{lunansky2022intervening}, we conceptualize the node importance evaluated by such methods as \textit{projected node importance}.} In early applications, researchers typically forced the value of a given node to 0 to simulate a state in which a symptom was completely eliminated in reality. They then compared changes in the total network score before and after the simulated intervention to determine which symptom, if eliminated, was projected to produce the greatest overall improvement in the network \parencite{castro2019differential,henry2022control,robinaugh2016identifying}. Nevertheless, from the perspective of clinical realism, such an absolute form of simulated manipulation has clear limitations \parencite{barth2016comparative}.

In response to this issue, \textcite{lunansky2022intervening} proposed the NodeIdentifyR Algorithm (NIRA) on the basis of previous work. This algorithm is built on the Ising model, which is specifically designed for binary variables. In the context of psychopathology, the Ising model simplifies symptoms into two states, presence or absence, and uses node intercepts to represent the spontaneous activation probability of each symptom. Unlike earlier approaches that forced a node to 0, NIRA simulates intervention effects by systematically adjusting the \textit{intercept} of each node. Decreasing or increasing the intercept corresponds, respectively, to a decrease or increase in the spontaneous activation probability of the symptom. This representation is more closely aligned with how interventions operate in clinical practice. In its implementation, the algorithm generates 5,000 simulated observations under each parameter adjustment condition and compares the post-intervention total network score with the untreated baseline result. The node that produces the largest change in the total score is ultimately identified as the symptom projected to be the strongest intervention target.

However, NIRA still has two limitations. First, this algorithm can currently operate only within the Ising model framework, which is designed for binary variables. This limitation substantially limits its applicability because the Gaussian graphical model is one of the most widely used statistical models in psychological network research involving continuous variables \parencite{burger2023reporting,epskamp2018tutorial,isvoranu2023estimation}. Consequently, NIRA cannot be directly applied to the large body of psychological network studies in which variables are modeled using a GGM. Second, the \textit{node intercept} manipulated by NIRA has a clear theoretical meaning in psychopathology. Once this parameter is taken outside that domain, however, it may be difficult to assign it an equally clear theoretical interpretation for psychological variables such as personality traits, emotional states, or attitudes and beliefs. This limitation constrains the extension of this approach to broader areas of psychology.

To address the above limitations, the present article proposes a Gaussian Graphical Model NodeIdentifyR Algorithm (GGMNIRA), also referred to as a Gaussian graphical model-based simulated manipulation algorithm. From a broader methodological perspective, GGMNIRA and the original NIRA can be understood as specific implementations of the same principle of simulated manipulation within different statistical models. Both approaches perturb the intercept in a node’s conditional model and examine the resulting projected changes. However, they differ in the interpretive emphasis placed on this perturbation. Whereas the original NIRA focuses primarily on changes in the intercept, GGMNIRA places greater emphasis on the corresponding change in the node’s conditional mean, rather than on the intercept parameter itself. More specifically, the conditional mean refers to the expected value of a node given the values of the other nodes in the network. Thus, for symptoms, personality traits, emotional states, or attitudes and beliefs that are appropriately modeled using a GGM, systematically changing a node’s conditional mean can be understood as changing the expected level of that variable given the values of the other variables in the model. Accordingly, the simulated manipulations considered in GGMNIRA are not restricted to psychopathology or clinical contexts but can also be applied to other areas of psychological research. Based on this broader model-level interpretation, throughout this article we prefer the term ``manipulation'' over ``intervention'' and primarily describe this procedure as the simulated manipulation of a node’s conditional mean. In what follows, we systematically introduce the GGMNIRA framework and provide an empirical example to illustrate its application in psychological network research. We further extend the GGMNIRA framework to \textit{bridge Gaussian graphical model}, thereby supporting its application to multi-construct networks. All of these methods are implemented in the R package \Rpkg{GGMNIRA}.

\section{Gaussian Graphical Model NodeIdentifyR Algorithm Framework}
\begin{figure}[ht]
    \centering
    \caption{Workflow of the GGMNIRA framework. After estimating the baseline Gaussian graphical model, each node's conditional mean is expressed as a weighted function of the other nodes. The simulated manipulation step is illustrated using \(X_2\) as an example; in the full procedure, each node is sequentially manipulated, and nodes are ranked by their KL divergence. The bottom panel shows the complementary procedures for stability assessment and difference testing.} 
    \label{fig:GGMNIRAworkflow}
    \apafiguregraphic{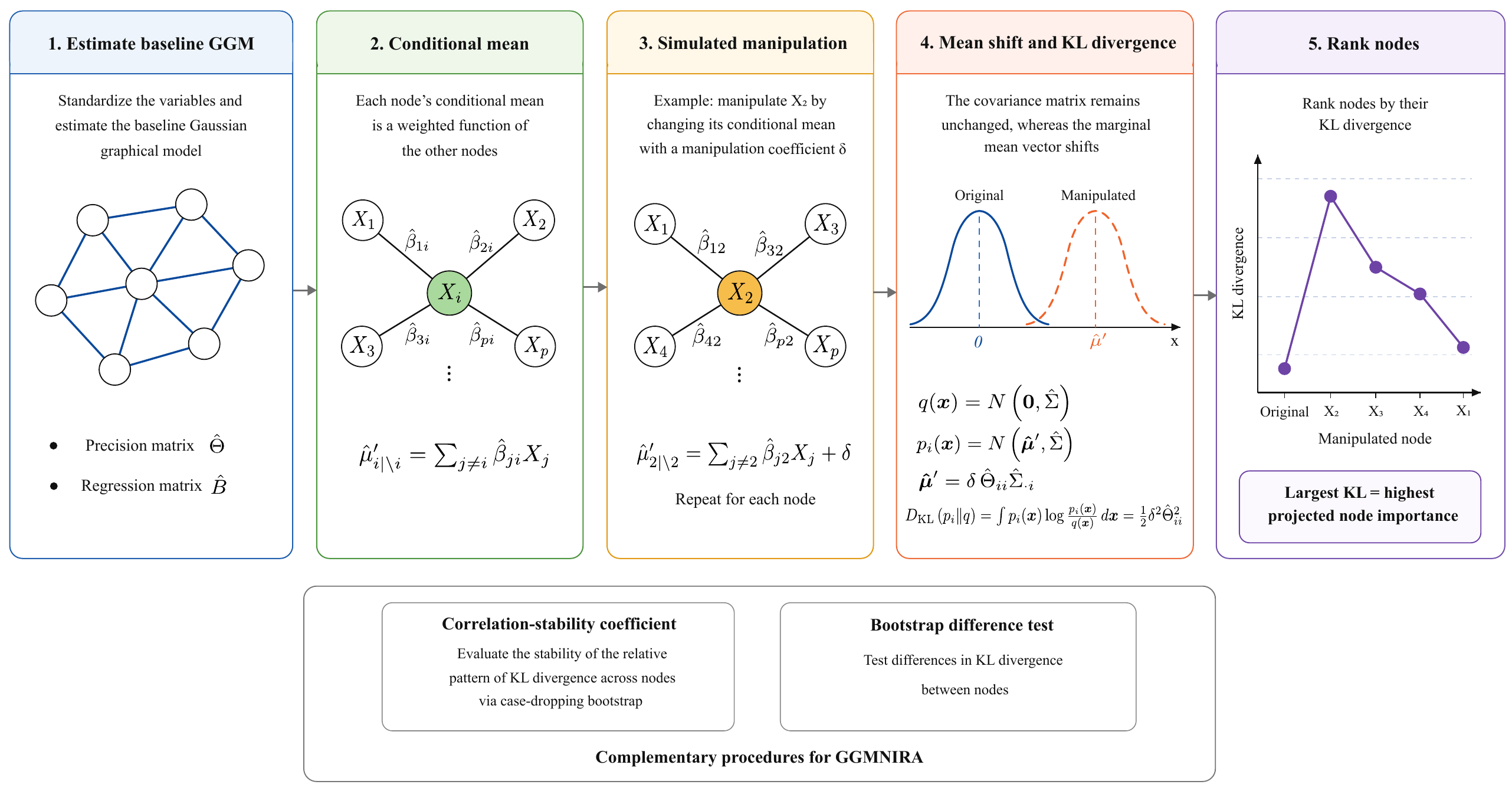}
\end{figure}

In this section, we systematically introduce GGMNIRA in three parts. First, we briefly review the construction of the Gaussian graphical model and its key properties, thereby providing the necessary foundation for the subsequent algorithmic derivation. We then focus on the core computational procedure of GGMNIRA. Finally, we introduce the complementary procedures for uncertainty assessment, including stability estimation and statistical inference. The corresponding workflow is shown in Figure~\ref{fig:GGMNIRAworkflow}.

\subsection{A Brief Review of Gaussian Graphical Model}
\label{sec:brief}
In this subsection, following the tutorial by \textcite{epskamp2018tutorial}, we briefly review the construction of Gaussian graphical model, further derive several important corollaries, and introduce the conditional Gaussian distribution. We hope that this review will facilitate readers' understanding of the present article. One point that warrants particular attention is that, from this subsection onward, all subsequent content in this article is described under the assumption that the variables have been standardized. We also repeatedly restate this important assumption in the key formula derivations below. This assumption is not arbitrary. Standardizing variables not only removes the influence of measurement scales for continuous variables and yields interpretable parameters \parencite{epskamp2018tutorial}, but also makes the strength of regularization comparable \parencite{carter2024partial}. Readers who are already familiar with this material may proceed directly to the second subsection.

Let us consider a $p$ dimensional random vector $\bm{X}^{T} = [X_1, X_2, ..., X_p]$, and let $\Sigma$ denote the variance-covariance matrix. Under the assumption that all variables have been standardized, we assume that $\bm{X}$ follows the following multivariate Gaussian distribution $q(x)$:
\[\bm{X} \sim N(0, \Sigma).\]\par
Compared with the variance-covariance matrix $\Sigma$, its inverse matrix, also known as the precision matrix, is more commonly used in Gaussian graphical model. We therefore denote it by $\Theta$:
\[\Theta = \Sigma^{-1}.\]\par
On the basis of the above multivariate Gaussian distribution, a Gaussian graphical model can be constructed, and the core of this construction lies in obtaining the partial correlation matrix. The off-diagonal elements of the partial correlation matrix are the partial correlation coefficients of primary interest to researchers. The magnitude and direction of a partial correlation coefficient reflect the strength and direction of the conditional association between variables. In particular, a partial correlation coefficient of zero indicates that the corresponding two variables are conditionally independent. There are two common approaches to obtaining partial correlation coefficients. The first approach is based on the precision matrix \parencite{lauritzen1996graphical}. After controlling for the remaining variables, the partial correlation coefficient between $X_i$ and $X_j$ can be expressed as
\begin{equation}
\text{Cor}(X_i, X_j \mid \bm{X}_{-(i,j)}) 
= - \frac{\Theta_{ij}}
        {\sqrt{\Theta_{ii\vphantom{j}}}\sqrt{\Theta_{jj}}}.
\label{eq:1} 
\end{equation}\par
The second approach uses \textit{node-wise regressions} \parencite{meinshausen2006high}. By constructing multiple regression models, the partial correlation coefficients can be obtained from the regression coefficients and regression residuals:
\begin{equation}
\text{Cor}(X_i, X_j \mid \bm{X}_{-(i,j)}) = \frac{\beta_{ji}\text{SD}(\varepsilon_j)}{\text{SD}(\varepsilon_i)} = \frac{\beta_{ij}\text{SD}(\varepsilon_i)}{\text{SD}(\varepsilon_j)}.
\label{eq:2} 
\end{equation}\par
Here, $\beta_{ji}$ denotes the regression coefficient of the predictor $X_j$ for the outcome variable $X_i$, after controlling for the remaining variables, and $\text{SD}(\varepsilon_j)$ denotes the standard deviation of the regression residuals when the outcome variable is $X_j$. In fact, the variance of the regression residuals corresponds to the diagonal elements of the precision matrix \parencite{pourahmadi2011covariance,epskamp2018gaussian}:
\begin{equation}
\text{Var}(\varepsilon_j) =\frac{1}{\Theta_{jj}}.
\label{eq:3} 
\end{equation}\par
Therefore, by combining Equations~\eqref{eq:1}, \eqref{eq:2}, and~\eqref{eq:3}, we obtain an important corollary:
\begin{equation}
\beta_{ji} = - \frac{\Theta_{ij}}{\Theta_{ii}}.
\label{eq:4}
\end{equation}\par
It is important to note that although $\beta_{ji}$ and $\beta_{ij}$ may be numerically close, they are not equal. This is because their numerators are numerically identical, whereas their denominators are not.

In addition, it is important to note that, in Gaussian graphical model, node-wise regressions is essentially formulated from the perspective of the conditional Gaussian distribution. This means that the mean and variance of the conditional distribution can also be expressed using graphical-model parameters and their derived parameters, such as $\Theta$ and $\beta$.

In a Gaussian graphical model consisting of $p$ continuous variables, for a given random variable $X_i$, conditional on $X_{\setminus i}=x_{\setminus i}$, the variable follows a conditional Gaussian distribution:
\begin{equation}
    P(X_i \mid x_{\setminus i}) = \frac{1}{\sqrt{2\pi \sigma_{i|\setminus i}^2}} \exp \left( - \frac{ \left( X_i - \mu_{i|\setminus i} \right)^2 }{ 2 \sigma_{i|\setminus i}^2 } \right).
    \label{eq:5}
\end{equation}\par
Here, $X \in \mathbb{R}^p$, $\mu_{i|\setminus i}$ is the conditional mean of variable $X_i$, and $\sigma_{i|\setminus i}^2$ is the conditional variance of variable $X_i$. After all variables have been standardized, the conditional mean and conditional variance are given by
\[\mu_{i|\setminus i} = \Sigma_{i, \setminus i} \Sigma_{\setminus i, \setminus i}^{-1} x_{\setminus i}= -\frac{\Theta_{i,\setminus i}}{{\Theta_{ii}}} x_{\setminus i},\]and
\[\sigma_{i|\setminus i}^2 = \Sigma_{ii} - \Sigma_{i, \setminus i} \Sigma_{\setminus i, \setminus i}^{-1} \Sigma_{\setminus i, i}= \frac{1}{{\Theta_{ii}}}.\]\par

According to the corollary in Equation~\eqref{eq:4}, the conditional mean can also be written as
\[\mu_{i|\setminus i} =\beta_{\setminus i,i} x_{\setminus i},\]
where $\beta_{\setminus i,i}$ is the vector of regression coefficients; that is, $\beta_{\setminus i,i} =[\beta_{1,i},\beta_{2,i}, ...,\beta_{p,i}]$. Through this transformation, we can see that the conditional mean is the sum of the products of the regression coefficients and the values of the corresponding random variables.

Regardless of the specific approach used to construct the graphical model, after obtaining the corresponding parameter expressions, the parameters generally need to be estimated through maximum likelihood estimation combined with a model selection algorithm, such as EBICglasso\footnote{Given its widespread use in estimating Gaussian graphical model, EBICglasso is adopted as the default model selection algorithm for constructing the GGM in the subsequent implementation of GGMNIRA.} \parencite{epskamp2018gaussian}. Therefore, for the conditional Gaussian distribution, the final parameter estimates can be expressed as
\[\hat\mu_{i|\setminus i} =\hat\beta_{\setminus i,i} x_{\setminus i},\]
and
\[\hat\sigma_{i|\setminus i}^2 = \frac{1}{{\hat\Theta_{ii}}}.\]

\subsection{Development of GGMNIRA}\par
In this subsection, we provide a detailed introduction to the construction process of GGMNIRA. In terms of the development logic, briefly, we simulate changes in the real world level of each node by sequentially manipulating its conditional mean. We then use the Kullback--Leibler divergence (KL divergence; \citeauthor{kullback1951kullback}, \citeyear{kullback1951kullback}) between the manipulated multivariate Gaussian probability distribution and the original baseline multivariate Gaussian probability distribution to quantify the global-level change in the network system. A larger KL divergence indicates a greater overall change in the network. This means that, if manipulating a particular node produces the largest KL divergence relative to the other nodes, then this node can be regarded as the node with the greatest \textit{projected importance} for the overall network. We provide a detailed mathematical derivation of this development logic in \nameref{sec:mathematical_derivation} subsection. In addition, to help readers gain a deeper understanding of the algorithm and apply it appropriately, we have added \nameref{sec:parameter-selection} subsection to explain why the conditional mean of a node was chosen as the manipulation parameter, and why KL divergence was chosen as the outcome measure.

\subsubsection{Parameter Selection}
\label{sec:parameter-selection}\par
The development of GGMNIRA requires two key methodological choices: what parameter in the Gaussian graphical model should be manipulated, and how the resulting network-level change should be quantified. In making these choices, both theoretical interpretability and statistical suitability need to be considered.

For Gaussian graphical model, the two available parameters are the conditional means of the nodes and the partial correlation matrix of the network. An important question, however, concerns the theoretical meaning of manipulating these two parameters. Addressing this question is essential because using a parameter solely because it is available, without first establishing its theoretical interpretation, may lead to serious theoretical errors.

In the field of psychopathology, the network theory of mental disorders proposed by \textcite{borsboom2017network} provides a corresponding theoretical framework. According to this theory, treating a mental disorder requires changing or manipulating the network itself, and such manipulations can be organized into three categories: \textit{symptom interventions}, which directly change the state of one or more symptoms; \textit{interventions in the external field}, which remove one or more triggering causes; and \textit{network interventions}, which change the network structure itself by modifying symptom-symptom connections. Within this psychopathology framework, manipulating the conditional mean of a node corresponds to changing the level of a symptom, whereas manipulating the partial correlation matrix corresponds to changing the associations among symptoms. This theoretical interpretation of the parameters is equally applicable to other domains of psychology. Therefore, if one aims to conduct simulated manipulation within a psychological network, the conditional means of the nodes and the partial correlation matrix of the network represent two theoretically meaningful candidate parameters for manipulation.

\begin{figure}[ht]
    \centering
    \caption{Direct manipulation on the marginal mean of a multivariate normal distribution}
    \label{fig:mnd}
    \apafiguregraphic{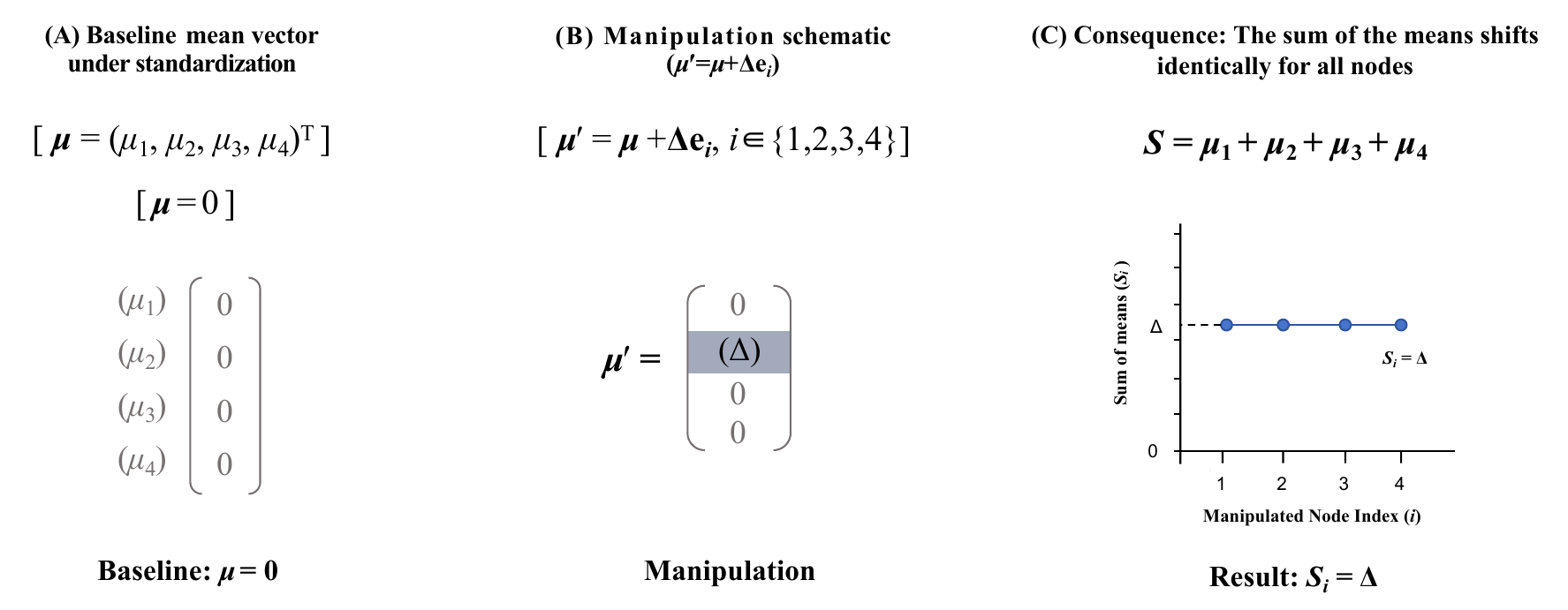}
\end{figure}

However, theoretical interpretability alone is not sufficient for selecting a manipulation parameter. If simulated manipulation is implemented merely in a mathematical or statistical sense, that is, by adding or subtracting a constant from the original parameter value, manipulating the partial correlation matrix is not practically viable. This is because one cannot ensure that, when the association between two nodes is changed, the associations among the remaining nodes will not also be affected in ways that cannot be captured. If this issue is ignored, such manipulation would remain purely theoretical and would have no practical use. By contrast, manipulating the conditional mean can address this problem statistically. The conditional mean of a node represents the average level of that node after controlling for the other nodes. When this average level changes, changes in the associations among other nodes can be allowed; such effects are, in essence, called \textit{moderation effects}, which we will briefly discuss in the \nameref{sec:dis} section. Therefore, we select the conditional mean of a node, rather than the partial correlation matrix, as the parameter for simulated manipulation. Readers may also wonder why the marginal mean vector of the multivariate Gaussian distribution is not selected as the parameter. Figure~\ref{fig:mnd} illustrates the potential problem associated with selecting this parameter: regardless of which node's marginal mean is changed, the sum of the means across all nodes changes to the same extent.

For the outcome measure used to quantify global-level changes in the network, we select KL divergence, which can effectively measure the discrepancy between the manipulated probability distribution and the original baseline probability distribution. We argue that, compared with the sum of the means, KL divergence can better reflect the underlying statistical nature.

\subsubsection{Mathematical Derivation of the Algorithm}
\label{sec:mathematical_derivation}\par
Building on the expressions for the estimated conditional mean and conditional variance in \nameref{sec:brief} section, after obtaining $\hat\mu_{i|\setminus i}$ and $\hat\sigma_{i|\setminus i}^2$ selected by EBICglasso, we implement simulated manipulation by adding a manipulation coefficient $\delta$ to $\hat\mu_{i|\setminus i}$ for the random variable $X_i$. The manipulated conditional mean, denoted by $\hat\mu_{i|\setminus i}'$, is given by
\begin{equation}
    \hat\mu_{i|\setminus i}' = \sum_{j \neq i} \hat\beta_{ji} X_j+\delta.
    \label{eq:6}
\end{equation}\par\footnote{From the perspective of node-wise regressions, Equation~\eqref{eq:6} shifts the node-specific intercept from 0 to $\delta$, because the baseline intercept is zero after standardization. We describe this same operation as a conditional-mean manipulation because, for any fixed values of the other nodes, it shifts the expected value of $X_i$ by $\delta$.}
Here, $\delta$ is a constant which expressed in standard-deviation units and has both magnitude\footnote{The magnitude of the manipulation coefficient is also referred to as manipulation intensity.} and direction. A larger absolute value of $\delta$ indicates a stronger manipulation; $\delta>0$ indicates an aggravating manipulation, whereas $\delta<0$ indicates an alleviating manipulation.

Because we aim to examine the overall change in the entire network following the manipulation, rather than only the local change in the manipulated variable itself, it is necessary to further obtain the manipulated marginal mean of the variable. By the law of total expectation, taking expectations on both sides of Equation~\eqref{eq:6} yields:
\begin{equation}
    E(\hat\mu_{i|\setminus i}') = \hat\mu_{i}'=\sum_{j \neq i} \hat\beta_{ji} \hat\mu_j'+\delta.
    \label{eq:7}
\end{equation}\par
Here, $\hat\mu_{i}'$ is the estimated marginal mean of the random variable $X_i$ after manipulation. For a more intuitive representation, Equation~\eqref{eq:7} can also be written in matrix form. Let $\boldsymbol{\hat\mu}'$ denote a $p\times 1$ vector composed of the estimated marginal means of the $p$ random variables after manipulation. Let $\hat{B}$ denote a regression coefficient matrix with zeros on the diagonal and satisfying $\hat{B}_{ij}= \hat\beta_{ji}$ for $i\neq j$. Let $\boldsymbol{\delta}$ denote a $p\times 1$ manipulation vector, with $\boldsymbol{\delta} = \delta \cdot {e}_i$, where ${e}_i$ is a vector whose $i$th element is 1 and whose remaining elements are 0. This means that, for the sequence of random variables not being manipulated, $X_{\setminus i}$, the corresponding manipulation terms are 0. The matrix form can then be written as
\begin{equation}
\boldsymbol{\hat\mu}'=\hat{B}\boldsymbol{\hat\mu}'+\boldsymbol{\delta}.
\label{eq:8}
\end{equation}\par
Rearranging Equation~\eqref{eq:8} yields
\begin{equation}
\boldsymbol{\hat\mu}' = (I - \hat{B})^{-1} \boldsymbol{\delta}.
\label{eq:9}
\end{equation}\par
It should be noted that Equation~\eqref{eq:9} requires the matrix ${(I - \hat{B})}$ to be invertible, or equivalently, $1\notin\sigma(\hat{B})$, where $\sigma(\hat{B})$ denotes the spectrum of $\hat{B}$.\par

Furthermore, using the relationship between the regression coefficients and the precision matrix given in Equation~\eqref{eq:4}, together with the fact that the covariance matrix is the inverse of the precision matrix, Equation~\eqref{eq:9} can be further expressed as
\begin{equation}
\boldsymbol{\hat\mu}'= \delta\,\hat\Theta_{ii}\hat\Sigma_{\cdot i},
\label{eq:10}
\end{equation}
where $\hat\Sigma$ denotes the covariance matrix obtained by inverting the estimated precision matrix $\hat\Theta$, and $\hat\Sigma_{\cdot i}$ denotes the $i$th column of this covariance matrix.

Equation~\eqref{eq:10} shows that, when a simulated manipulation is applied to the conditional mean of a given random variable,  the manipulated mean vector has a closed-form analytical solution.

Building on this result, a further question concerns whether the covariance matrix of the network changes following the manipulation. Under the current Gaussian graphical model framework, applying a constant shift either to a random variable itself or to its conditional mean changes only the mean levels of the variables, while their variances and covariances with other variables remain unchanged. We therefore consider that the covariance structure among the variables in the network to remain invariant under the manipulation. 

Given that the manipulation changes the mean vector while leaving the covariance matrix unchanged, the manipulated joint distribution can therefore be directly obtained. Specifically, after manipulating the conditional mean of random variable $X_i$, the corresponding multivariate Gaussian distribution can be written as $p_i(\boldsymbol{x})=
N\left(
\delta\hat{\Theta}_{ii}\hat{\Sigma}_{\cdot i},
\hat{\Sigma}
\right),$
whereas the baseline distribution is $
q(\boldsymbol{x})
=
N\left(
\boldsymbol{0},
\hat{\Sigma}
\right).$
We can then quantify the global-level change in the network by computing the KL divergence between the manipulated distribution $p_i(\boldsymbol{x})$ and the baseline distribution $q(\boldsymbol{x})$. Because the two distributions have the same covariance matrix, the KL divergence can be expressed as
\begin{equation}
\begin{aligned}
D_{\mathrm{KL}}\left(p_i \middle\| q\right)
&=
\int
p_i(\boldsymbol{x})
\log
\frac{p_i(\boldsymbol{x})}
{q(\boldsymbol{x})}
\,d\boldsymbol{x}
\\
&=
\frac{1}{2}
\boldsymbol{\hat{\mu}}'^{\top}
\hat{\Theta}
\boldsymbol{\hat{\mu}}'
\\
&=
\frac{1}{2}
\delta^2
\hat{\Theta}_{ii}^{2}
\hat{\Sigma}_{\cdot i}^{\top}
\hat{\Theta}
\hat{\Sigma}_{\cdot i}
\\
&=
\frac{1}{2}
\delta^2
\hat{\Theta}_{ii}^{2}
\hat{\Sigma}_{ii}
\\
&=
\frac{1}{2}
\delta^2
\hat{\Theta}_{ii}^{2}.
\end{aligned}
\label{eq:11}
\end{equation}\par
Next, we only need to sequentially manipulate the conditional mean of each random variable and compute the corresponding KL divergence according to Equation~\eqref{eq:11}. Finally, all variables are ranked in descending order according to their KL divergence values. The variable with the largest KL divergence is the variable with the greatest projected importance in network.\footnote{In the Simulation Study section, we further conduct a recovery simulation to evaluate the performance of GGMNIRA.} In other words, manipulating this variable is projected to have the largest impact on the network.

However, implementing the above algorithm still requires specifying the manipulation coefficient $\delta$. It is therefore necessary to further consider how the magnitude and direction of the manipulation should be determined. Unlike original NIRA, in which the projected importance ranking of variables may change when the manipulation intensity or manipulation direction is altered, Equation~\eqref{eq:11} shows that the projected importance ranking of variables does not change regardless of the values assigned to the manipulation intensity and manipulation direction. The manipulation intensity only affects the magnitude of the manipulation effect, whereas changing the manipulation direction has no effect on the KL divergence. We argue that this difference fundamentally arises from the distinct model assumptions underlying the two approaches, with GGMNIRA relying on a linear system and the original NIRA relying on a nonlinear system. In addition, although changes in manipulation intensity do not alter the projected importance ranking of variables, as noted above, determining the minimum manipulation intensity required for the manipulation to be effective remains meaningful from a practical perspective. We therefore determine this value through the following derivation.

In multivariate statistics, the Mahalanobis distance $D_{M}$ can be used to quantify the standardized distance between a vector and the center of a distribution \parencite{mahalanobis2018generalized}. In the context of the present subsection, the KL divergence in Equation~\eqref{eq:11} is numerically equivalent to one half of the squared Mahalanobis distance between the manipulated mean vector $\boldsymbol{\hat\mu}'$ and the center of the baseline distribution; that is,
\begin{equation}
\begin{aligned}
D_{KL}(p_i \| q) =  \frac{1}{2} D_{M}^2.
\end{aligned}
\label{eq:12}
\end{equation}

This means that, with the covariance matrix held constant, KL divergence quantifies the distance between the center of the manipulated distribution and the center of the baseline distribution, or the magnitude of the mean shift. Furthermore, based on the theory of probability ellipsoids for the multivariate normal distribution, we know that, for a random vector following a $p$-dimensional multivariate normal distribution, its squared Mahalanobis distance from the distribution mean follows a chi-square distribution with $p$ degrees of freedom, denoted by $\chi_p^2$ \parencite{johnson2002applied}. This implies that exactly $(1-\alpha)$ of the probability mass is contained within the ellipsoid centered at the mean, whose boundary is defined by $\chi^2_{p,\,1-\alpha}$. Based on the equivalence in Equation~\eqref{eq:12}, when the KL divergence between the manipulated distribution and the baseline distribution satisfies
\[
2D_{KL}(p_i \| q) > \chi^2_{p,\,0.95},
\]
the mean-shift vector induced by the manipulation falls outside the boundary of the 95\% probability ellipsoid of the baseline distribution. In other words, the manipulation effect exceeds the range of natural variation covering the vast majority of observations under the baseline distribution. We regard this as a criterion for determining whether the manipulation is effective. It should be emphasized that this criterion concerns the distributional shift from a geometric perspective and is not a conventional hypothesis test. Combining this criterion with Equation~\eqref{eq:11}, we can further derive the manipulation intensity required for the manipulation effect on the random variable $X_i$ to just exceed the above 95\% probability ellipsoid boundary:
\begin{equation}
\begin{aligned}
|\delta_{i}| > \frac{\sqrt{\chi^2_{p,0.95}}}{\hat{\Theta}_{ii}}.
\end{aligned}
\label{eq:13}
\end{equation}

Equation~\eqref{eq:13} shows that, for different variables within the same network, the minimum manipulation intensity required for the manipulation to be effective is not identical. To ensure concise output, we adopt a global minimum manipulation intensity strategy. Specifically, we take the smallest value among the minimum manipulation intensities required across all variables in the network and round it upward, using the resulting value as the unified manipulation intensity in practical applications. Consider a three-variable network as an example. If the minimum manipulation intensities required for variables $A$, $B$, and $C$ are $1.5$, $2.6$, and $0.8$, respectively, we take the smallest of these values, $0.8$, and round it upward to obtain $\delta = 1$, which is then used as the final unified manipulation intensity. In addition, it should be noted that the threshold for evaluating the distributional shift differs across networks with different numbers of variables, which readers should keep in mind.

After determining how $\delta$ should be specified, the main framework of GGMNIRA is complete. Building on this framework, we can further examine the mathematical nature of GGMNIRA and, in doing so, clarify its relationship with some existing measures commonly used in psychological network. Equation~\eqref{eq:11} shows that, when the manipulation coefficient $\delta$ is held fixed, the KL divergence is entirely determined by the corresponding diagonal element of the estimated precision matrix, $\hat\Theta_{ii}$. The quantity $\hat\Theta_{ii}$ can itself be viewed as a statistic that incorporates information from both direct and indirect pathways throughout the network. From this perspective, it is reasonable to interpret the KL divergence as a measure of the overall change in the network. This observation also raises a further question: given that the KL divergence is determined by the parameters of the graphical model, might it also be related to existing measures commonly used in network analysis, such as \textit{strength centrality}?

Strength centrality is constructed from the partial correlation structure of the network, and, as shown in Equation~\eqref{eq:1}, the precision matrix determines this partial correlation structure. It is therefore plausible that GGMNIRA result is related to strength centrality. We provide a detailed discussion of this relationship in the online supplementary materials.\footnote{The online supplementary materials are available at \url{https://github.com/ywu6888/Supplementary-materials-for-GGMNIRA}.} Here, however, we focus on another measure that has a more direct mathematical relationship with GGMNIRA, that is, \textit{predictability} proposed by \textcite{haslbeck2018well}. For continuous variables, \textcite{haslbeck2018well} quantified predictability using the proportion of explained variance,
\begin{equation}
R_i^2
=
1-
\frac{\operatorname{Var}(\hat{X}_i-X_i)}
{\operatorname{Var}(X_i)},
\label{eq:14}
\end{equation}
where $\hat{X}_i$ denotes the values of $X_i$ predicted from the other nodes in the fitted network model. Thus, $R_i^2$ indicates the extent to which the variance of node $X_i$ can be explained by the other nodes in the network.

This measure was originally operationalized by comparing model-based predictions with the observed values. Within a Gaussian graphical model, however, the corresponding predictability can also be derived directly from the estimated model parameters. Combining the definition of explained variance with the relationship between conditional variance and the precision matrix given in Equation~\eqref{eq:3}, we refer to the resulting measure as \textit{GGM-based predictability}, which can be expressed as
\begin{equation}
\hat{R}_i^2
=
1-\frac{1}{\hat{\Theta}_{ii}}.
\label{eq:15}
\end{equation}\par
Equation~\eqref{eq:15} shows that GGM-based predictability, like the KL divergence in Equation~\eqref{eq:11}, is completely determined by the diagonal element $\hat{\Theta}_{ii}$ of the estimated precision matrix. This common dependence makes it possible to derive an explicit mathematical relationship between the two measures. Specifically, rearranging Equation~\eqref{eq:15} in terms of $\hat{\Theta}_{ii}$ gives
\[
\hat{\Theta}_{ii}
=
\frac{1}{1-\hat{R}_i^2}.
\]

Substituting this expression into Equation~\eqref{eq:11} yields
\begin{equation}
D_{\mathrm{KL}}(p_i\|q)
=
\frac{\delta^2}
{2(1-\hat{R}_i^2)^2}.
\label{eq:16}
\end{equation}\par
Equation~\eqref{eq:16} shows that, for a fixed manipulation coefficient, the KL divergence in GGMNIRA is a strictly increasing function of GGM-based predictability. Therefore, the two measures yield the same ranking of nodes. This correspondence is not unexpected, because both measures are derived within the same Gaussian graphical model and are ultimately determined by the same underlying model parameters. Given the relatively constrained parameterization of the GGM, exact mathematical relationships between different node-level measures can naturally arise.

Importantly, despite this mathematical correspondence, the two measures are based on fundamentally different perspectives. Predictability describes the extent to which a node can be predicted by the other nodes in the network, whereas GGMNIRA starts from the perspective of simulated manipulation and evaluates projected node importance by quantifying the projected change in the overall network distribution induced by that manipulation. Thus, the correspondence established above concerns the results obtained under the current GGM framework and does not imply equivalence in their conceptual interpretations.

Moreover, we consider that GGMNIRA should be viewed as an underlying framework for simulated manipulation rather than merely as a single node-level measure. On the basis of this framework, complementary procedures like stability estimation and statistical inference can be developed, as introduced in the next subsection. Importantly, this framework can also be extended to more complex network settings, including bridge Gaussian graphical model, which will be discussed later in this paper, as well as moderated and temporal Gaussian graphical model. Above all, we argue that GGMNIRA retains meaningful methodological and theoretical value within the broader framework of simulated manipulation in psychological networks.

\subsection{Stability Estimation and Statistical Inference}
As noted above, GGMNIRA can be further supplemented by complementary procedures. In this section, we focus on two of the most important ones: stability estimation and statistical inference. These procedures are necessary because KL divergence is computed from an estimated network structure, and sampling variability may affect both the relative pattern of KL divergence across nodes and the estimated differences between specific nodes. Therefore, without further assessment of uncertainty, directly interpreting the node with the largest estimated KL divergence as having the greatest projected importance, or treating observed between-node differences as substantively meaningful, may lead to misleading conclusions.

Specifically, we use the correlation-stability coefficient to evaluate the stability of the relative pattern of KL divergence across nodes under case-dropping bootstrap resampling, and the bootstrap difference test to examine whether the KL divergence values of two nodes differ significantly. The following subsections describe these two procedures in detail.

\subsubsection{Stability Estimation}
\begin{figure}[!htbp]
    \centering
    \caption{Illustration of sampling-induced uncertainty in KL divergence estimates. \textit{Panel A} shows the population eight-node ring network with positive partial correlations of 0.25, and \textit{Panel B} shows the corresponding population KL divergences, which are identical across nodes ($KL \approx 0.66$). \textit{Panel C} shows the network estimated from a single sample of $n = 500$ using EBICglasso ($\gamma = 0.50$), and \textit{Panel D} shows the corresponding estimated KL divergences, which vary across nodes despite their identical population values (ranging from approximately 0.54 to 0.65).}
\label{fig:eac}
    \apafiguregraphic{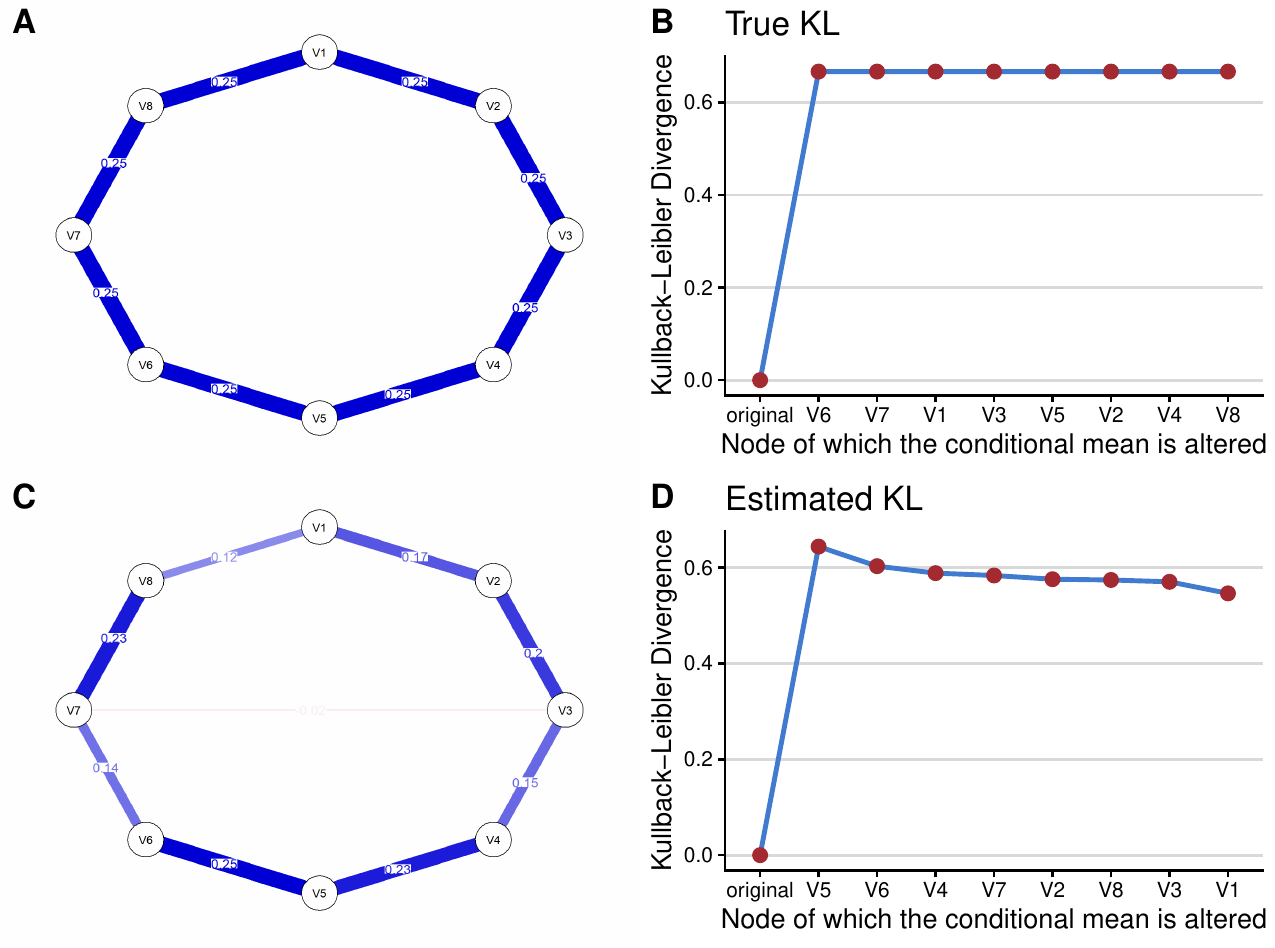}
\end{figure}

As with other statistical parameters, KL divergence may be incorrectly estimated because of uncertainty arising from empirical sampling, which may in turn lead to erroneous conclusions. To illustrate this issue more intuitively, we followed the study by \textcite{epskamp2018estimating} and constructed the simulated network structure shown in Figure~\ref{fig:eac}A. This ring-shaped Gaussian graphical network consists of eight nodes, with each node connected only to its two adjacent nodes, and the edges represent partial correlation coefficients. In addition, all edges in the network are positive, with their magnitudes set to 0.25. We then applied GGMNIRA to separately perform simulated manipulation on each node in the network, with the results shown in Figure~\ref{fig:eac}B. The KL divergences were identical across all nodes, indicating that, under this network structure, each node has equal projected importance in the network. Subsequently, we treated this network as the true network and generated 500 simulated observations from the corresponding multivariate normal distribution. We then re-estimated the network structure based on this sample, with the results shown in Figure~\ref{fig:eac}C. As can be seen, the estimated network structure was relatively similar to the true network, but certain discrepancies remained, particularly the emergence of unexpected weak negative edges. Similarly, we applied GGMNIRA to separately perform simulated manipulation on each node in the network, with the results shown in Figure~\ref{fig:eac}D. At this point, differences emerged in the KL divergences across nodes. If the accuracy of the estimation cannot be established, researchers may draw an erroneous conclusion from these results, namely, that node V5 has greater projected importance than the other nodes.

In general, when evaluating estimation accuracy, nonparametric bootstrap methods are commonly used to construct confidence intervals for statistical quantities \parencite{efron1992bootstrap}. However, directly applying this approach to KL divergence still involves several methodological limitations. First, as shown in Equation~\eqref{eq:11}, KL divergence is a quadratic function of the estimated diagonal element of the precision matrix, $\hat{\Theta}_{ii}$. Because $\hat{\Theta}$ is obtained after regularization and model selection, its estimates may contain shrinkage and selection bias, which is inherited by the KL divergence and further transformed through the quadratic relationship. Second, the sparse network structure selected by EBICglasso may vary across bootstrap samples, particularly for edges whose estimated values are close to zero \parencite{potscher2009distribution,williams2021confidence}. Such changes in the selected network structure also affect the estimated precision matrix and its diagonal elements, resulting in irregular variation in the bootstrap distribution of KL divergence. Consequently, percentile bootstrap confidence intervals may not provide reliable coverage for KL divergence.

\begin{figure}[!htbp]
    \centering
    \caption{Bootstrap 95\% CI of the KL divergence; Red dot = sample estimate, gray line = Bootstrap 95\% CI, blue line = true value.}
    \label{fig:eaci}
    \apafiguregraphic{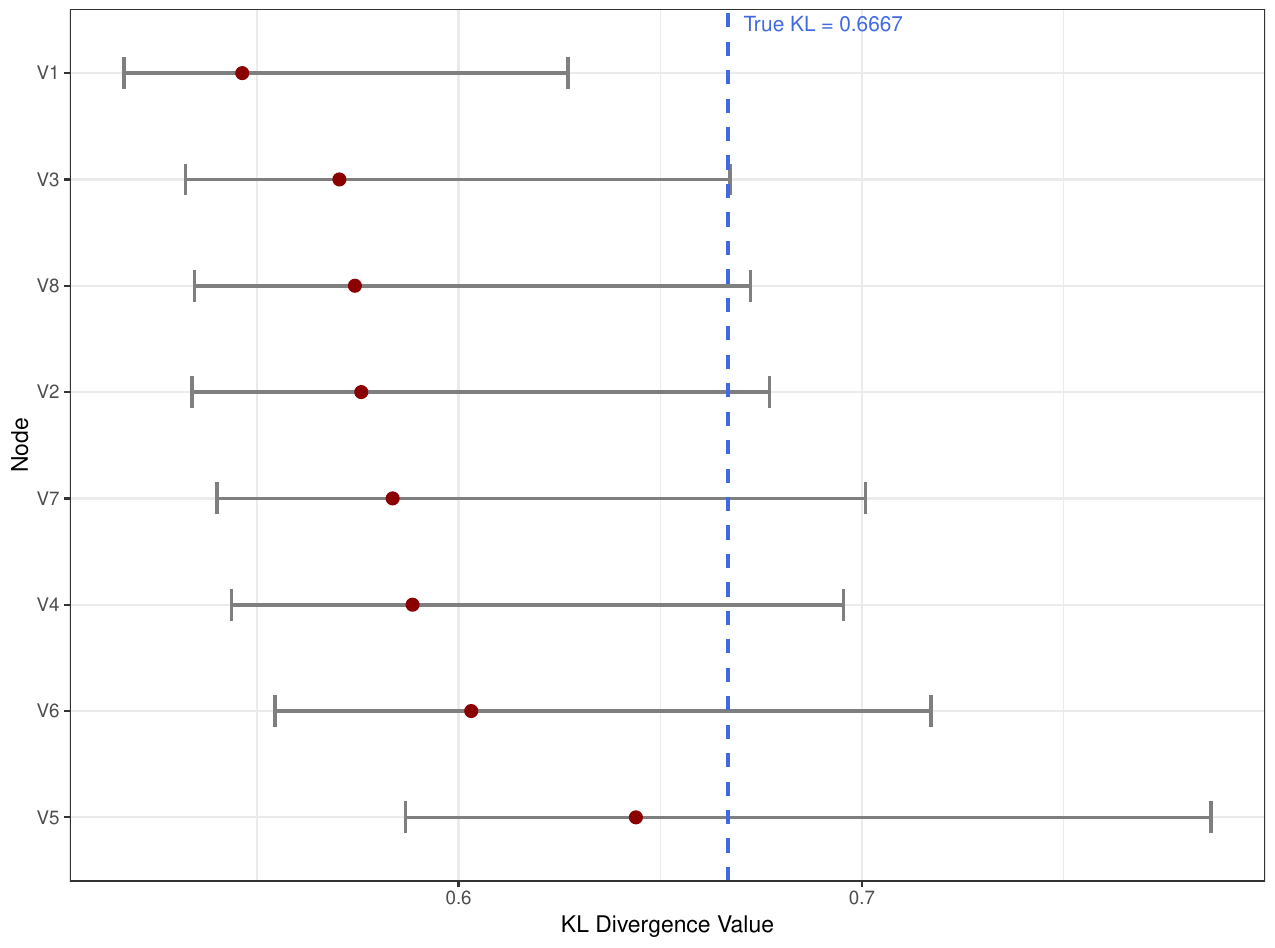}
\end{figure}

To examine whether the above limitations arise in practice, we conducted 1,000 bootstrap resamples based on the 500 simulated observations generated above to construct 95\% confidence intervals for the KL divergence. This procedure was intended to mimic the practical difficulty researchers would face if they used bootstrap methods to assess the accuracy of KL divergence in empirical analyses. As shown in Figure~\ref{fig:eaci}, the 95\% confidence intervals for some nodes, such as V1 and V3, failed to include or only marginally covered the true KL divergence, confirming the presence of systematic bias in these confidence intervals. In addition, the widths of the confidence intervals differed substantially across nodes, further indicating that the randomness of the LASSO sparsity pattern seriously interfered with the estimation results. Based on these findings, we argue that using bootstrap confidence intervals to evaluate the accuracy of KL divergence estimation is unreliable. As an alternative, we suggest that the \textit{correlation stability coefficient}, or \textit{CS coefficient}, proposed by \textcite{epskamp2018estimating}, can be used for stability estimation. 

The CS coefficient evaluates stability by examining whether the relative pattern of KL divergence across nodes remains consistent as the number of observations is gradually reduced. Specifically, this method adopts a \textit{case-dropping subset bootstrap} framework: subsets containing different proportions of the original observations are repeatedly drawn, the network is re-estimated in each subset, and the corresponding KL divergence values are then computed. The KL divergence vector obtained from the full sample is used as the reference, and its Pearson correlation with the KL divergence vector obtained from each subset is calculated to quantify the stability of the relative pattern across nodes. If this correlation decreases substantially after only a small proportion of observations has been removed, the relative pattern of KL divergence is highly sensitive to sampling variability. In such cases, substantive interpretations of differences between nodes would face a considerable risk of estimation error.

On this basis, and following the definition of the CS coefficient, we define the $CS(cor=0.70)$ coefficient for KL divergence as the maximum proportion of observations that can be dropped while ensuring that at least 95\% of the bootstrap subsamples have a Pearson correlation coefficient of no less than 0.70 with the KL divergence obtained from the original sample. The value of 0.70 is used as the default threshold for the correlation coefficient because it is commonly regarded as a benchmark for a large effect size in the behavioral sciences \parencite{cohen2013statistical}. In addition, in the actual computation of the CS coefficient, if a bootstrap iteration fails to produce valid KL divergence because of estimation failure, such as violation of the invertibility condition, that resample is not included among the valid bootstrap samples. Instead, we adopt a discard-and-retry strategy until the prespecified number of valid case-dropping bootstrap samples is reached or the maximum number of attempts has been exhausted.

It should be emphasized that the empirical thresholds for the CS coefficient proposed by \textcite{epskamp2018estimating} for traditional centrality indices, namely, not lower than 0.25 and preferably higher than 0.50, may not be applicable to KL divergence because KL divergence and traditional centrality indices have different mathematical properties. Therefore, the original empirical thresholds should not be assumed to transfer directly to KL divergence.

Given these considerations, we further conducted a simulation study, reported in detail in the \nameref{sec:ss} section, to preliminarily establish empirical reference thresholds for the CS coefficient of KL divergence. The results showed that, when the CS coefficient for KL divergence is below 0.25, we do not recommend that researchers provide substantive interpretations of differences in KL divergence between nodes, as such results may lack sufficient stability. For reliable interpretation of between-node differences in KL divergence, the ideal threshold for the CS coefficient needs to be higher than 0.40. It should be noted that these thresholds were established based on simulation results and should not be treated as rigid guidelines.

\subsubsection{Bootstrap Difference Test}
Testing whether there are significant differences in KL divergence between nodes is another procedure that warrants further development, as it may help researchers more accurately identify the node with the greatest projected importance in the network. This test can be developed by referring to the bootstrap confidence interval approach for differences in node centrality proposed by \textcite{epskamp2018estimating}. In the present study, the KL divergence obtained from nonparametric bootstrap resampling can be used to construct confidence intervals for pairwise node differences, thereby testing whether the KL divergences of two nodes differ significantly. Specifically, for each pair of nodes $(i,j)$, the difference $d^{(b)} = KL_j^{(b)} - KL_i^{(b)}$ is computed in each bootstrap iteration. Based on the sequence of differences obtained from $N$ bootstrap iterations, the $\alpha/2$ and $1-\alpha/2$ quantiles \parencite{hyndman1996sample} are used to construct a 95\% confidence interval. If this interval does not include 0, the null hypothesis that the two nodes have equal KL divergence is rejected. 

However, as discussed in the previous subsection, one unresolved challenge remains: because of the intrinsic properties of KL divergence, directly using bootstrap methods to construct confidence intervals may produce substantial bias. Thus, an important question is whether confidence intervals for between-node differences in KL divergence constructed using this approach remain reliable. To address this question, we conducted an additional simulation study, also reported in detail in the \nameref{sec:ss} section, to preliminarily evaluate the validity of applying this difference testing procedure to KL divergence.

The results indicated that this test is approximately valid for KL divergence. Specifically, except for the highly conservative result at $n=100$, the observed Type I error rates were close to the nominal significance level of $\alpha=.05$. Statistical power also increased with sample size and greater population KL divergence heterogeneity. However, achieving relatively high power, such as 0.80, still required a large sample size of 5,000, which is uncommon in most psychological studies. Therefore, this finding should be understood as indicating that, in most psychological research contexts, the power of the bootstrap difference test remains limited. Under such circumstances, researchers should not routinely interpret nonsignificant results as supportive evidence that the KL divergence values of two nodes are equal \parencite{wagenmakers2007practical}; rather, they should state only that there is no significant difference in KL divergence between the nodes. In addition, given the properties of the KL divergence, we argue that correction for multiple comparisons should still not be applied to the bootstrap difference test at this stage. Readers may refer to \textcite{epskamp2018estimating} for the specific reasons.

\section{Simulation Study}
\label{sec:ss}\par
To systematically evaluate the performance of GGMNIRA and its complementary statistical procedures, we conducted three simulation studies. Specifically, the first simulation study examined whether GGMNIRA results estimated from sample data could recover the population-level pattern of KL divergence and identify the projected most important node under different sample sizes and levels of population KL divergence heterogeneity. The second simulation study primarily aimed to establish preliminary empirical reference thresholds for the CS coefficient of KL divergence, while also examining its performance across different simulation conditions. The third simulation study assessed the statistical validity of applying the bootstrap difference test to KL divergence. 

All three simulation studies followed a common design that varied sample size and population KL divergence heterogeneity and used the same framework for generating and screening population networks.  First, for population KL divergence heterogeneity, we operationalized this factor at four levels: \textit{Equal}, \textit{Low}, \textit{Medium}, and \textit{High}. To construct population networks corresponding to these four levels, we adopted the following generation and screening procedure. 

We used an eight-node Gaussian graphical model with a ring topology as the base network, in which each node was connected to its two adjacent nodes and all partial correlations were fixed at 0.25. Candidate population networks were generated from this base network using a random rewiring procedure\parencite{watts1998collective}. A rewiring probability of 0 corresponded to the original ring network, whereas for rewired candidates the proposal rewiring probability was randomly selected from 0.10, 0.50, and 1.00. Importantly, rewiring probability served only as a mechanism for generating candidate networks and was not treated as a formal simulation factor. Instead, the realized population KL divergence vector of each candidate network was used to determine its population KL divergence heterogeneity condition.

Specifically, a network was classified into the Equal condition when the range of its population KL divergence values was smaller than $10^{-6}$. For the Low, Medium, and High conditions, the difference between the largest and second-largest population KL divergence values was first required to be at least 0.1, ensuring that the projected most important node was sufficiently distinguishable from the second-ranked node. Population KL divergence heterogeneity was then quantified by the population standard deviation of the KL divergence values across nodes. Networks were classified as Low when $0.055 \leq SD_{\mathrm{KL}} < 0.075$, Medium when $0.095 \leq SD_{\mathrm{KL}} < 0.110$, and High when $SD_{\mathrm{KL}} \geq 0.150$. Candidate networks falling outside these predefined regions were excluded. These thresholds were determined through a preliminary candidate-pool calibration procedure described in the online supplementary materials.

For sample size, we likewise considered four levels: 100, 500, 1,000, and 5,000, resulting in a $4 \times 4$ simulation design crossing sample size with population KL divergence heterogeneity. For each combination of population KL divergence heterogeneity and sample size, 500 valid replications were independently obtained. Within each replication, simulated data were first generated from the multivariate normal distribution corresponding to the population network. The network structure was then re-estimated using the EBICglasso model-selection algorithm with $\gamma = 0.50$, and the corresponding KL divergence values were subsequently computed from the estimated network using GGMNIRA. The manipulation coefficient was fixed at 1 throughout the three simulation studies. 

\subsection{Simulation Study on Recovery}
Before evaluating the complementary uncertainty estimation procedure, we first examined the recovery performance of GGMNIRA itself under the sample sizes and population KL divergence heterogeneity conditions described above. 

For the Low, Medium, and High conditions, recovery was evaluated using three complementary indices. First, we calculated the Pearson correlation between the population and estimated KL divergence vectors to assess the recovery of the relative pattern of KL divergence across nodes. Second, we calculated the Top-1 recovery rate, defined as the proportion of replications in which the node with the largest estimated KL divergence was identical to the node with the largest KL divergence in the population network. Third, we evaluated the numerical estimation error of KL divergence using the normalized root mean square error (nRMSE). Lower nRMSE values indicate more accurate recovery of the population KL divergence values. For the Equal condition, because all nodes have identical population KL divergence values, the Pearson correlation and Top-1 recovery rate are not defined; therefore, only nRMSE was examined.

\begin{figure}[!htbp]
    \centering
    \caption{GGMNIRA recovery across sample sizes and levels of population KL divergence heterogeneity. Pearson correlation reflects recovery of the relative KL divergence pattern, Top-1 recovery reflects correct identification of the node with the largest population KL divergence, and nRMSE reflects numerical estimation error, with lower values indicating better recovery. The Equal condition is not shown because Pearson correlation and Top-1 recovery are undefined under this condition.}
    \label{fig:recovery}
    \apafiguregraphic{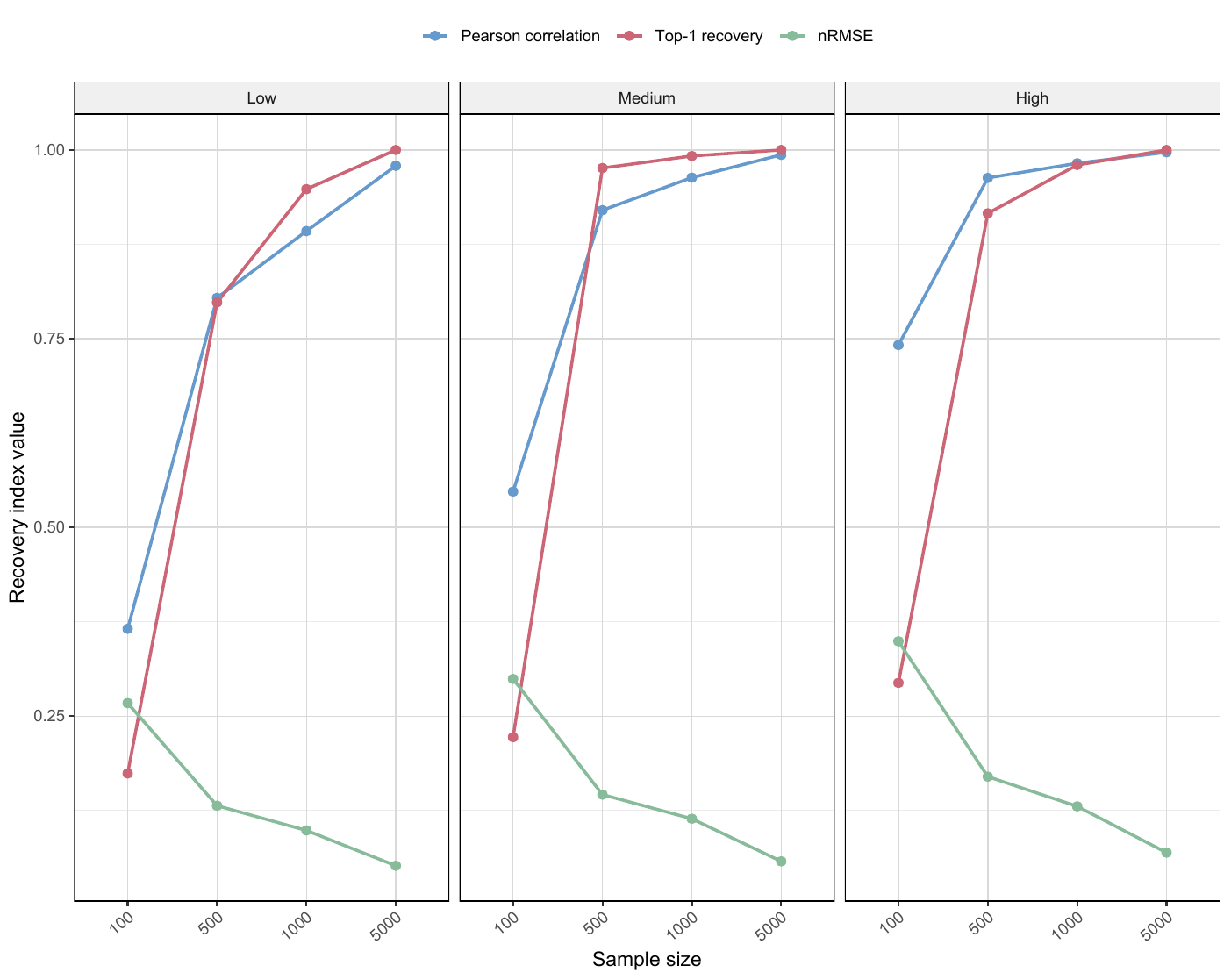}
\end{figure}

Figure~\ref{fig:recovery} presents the recovery results for the Low, Medium, and High conditions. The Pearson correlation between the estimated and population KL divergence vectors increased consistently with sample size across all three conditions, reaching approximately .98 or higher at $n=5{,}000$. Top-1 recovery showed a similar pattern. At $n=100$, the recovery rate was relatively low, ranging from .174 to .294, whereas it increased to .948 or higher at $n=1{,}000$ and reached 1.00 in all three conditions at $n=5{,}000$. The nRMSE also decreased consistently with increasing sample size, indicating progressively smaller numerical estimation error. Under the Equal condition, nRMSE decreased from .241 at $n=100$ to .048 at $n=5{,}000$. Overall, these results indicate that GGMNIRA recovery becomes markedly more accurate as sample size increases, whereas recovery remains limited at relatively small sample sizes.

\subsection{Simulation Study on CS Coefficient}
Having established the recovery performance of GGMNIRA, this simulation study primarily aimed to establish preliminary empirical reference thresholds for the CS coefficient of KL divergence, while also examining its performance across different simulation conditions. Following the general calibration logic of \textcite{epskamp2018estimating}, we used the distribution of CS coefficients under the Equal condition as the reference for establishing the thresholds.  Specifically, the 75th percentile of this distribution was used as the minimum threshold for indicating stability, whereas the 95th percentile was used as the preferred threshold under ideal conditions.

Within each simulation replication, after the KL divergence vector had been estimated from the simulated data as described above, this full-sample KL divergence vector was used as the reference pattern. We then conducted case-dropping bootstrap resampling to compute the CS coefficient for KL divergence. Specifically, 15 case-dropping levels were considered, with the proportion of cases removed ranging from 5\% to 75\%. A total of 1,000 case-dropping bootstrap samples were generated for each simulated data, with the dropping proportion randomly selected from these 15 levels in each bootstrap sample.

\begin{figure}[!htbp]
    \centering
    \caption{Distribution of KL-CS coefficients across sample sizes and levels of population KL divergence heterogeneity. KL-CS denotes the correlation-stability coefficient for KL divergence. The Equal condition was used to establish the preliminary reference thresholds of 0.25 and 0.40.}
    \label{fig:kl_cs_simulation}
    \apafiguregraphic{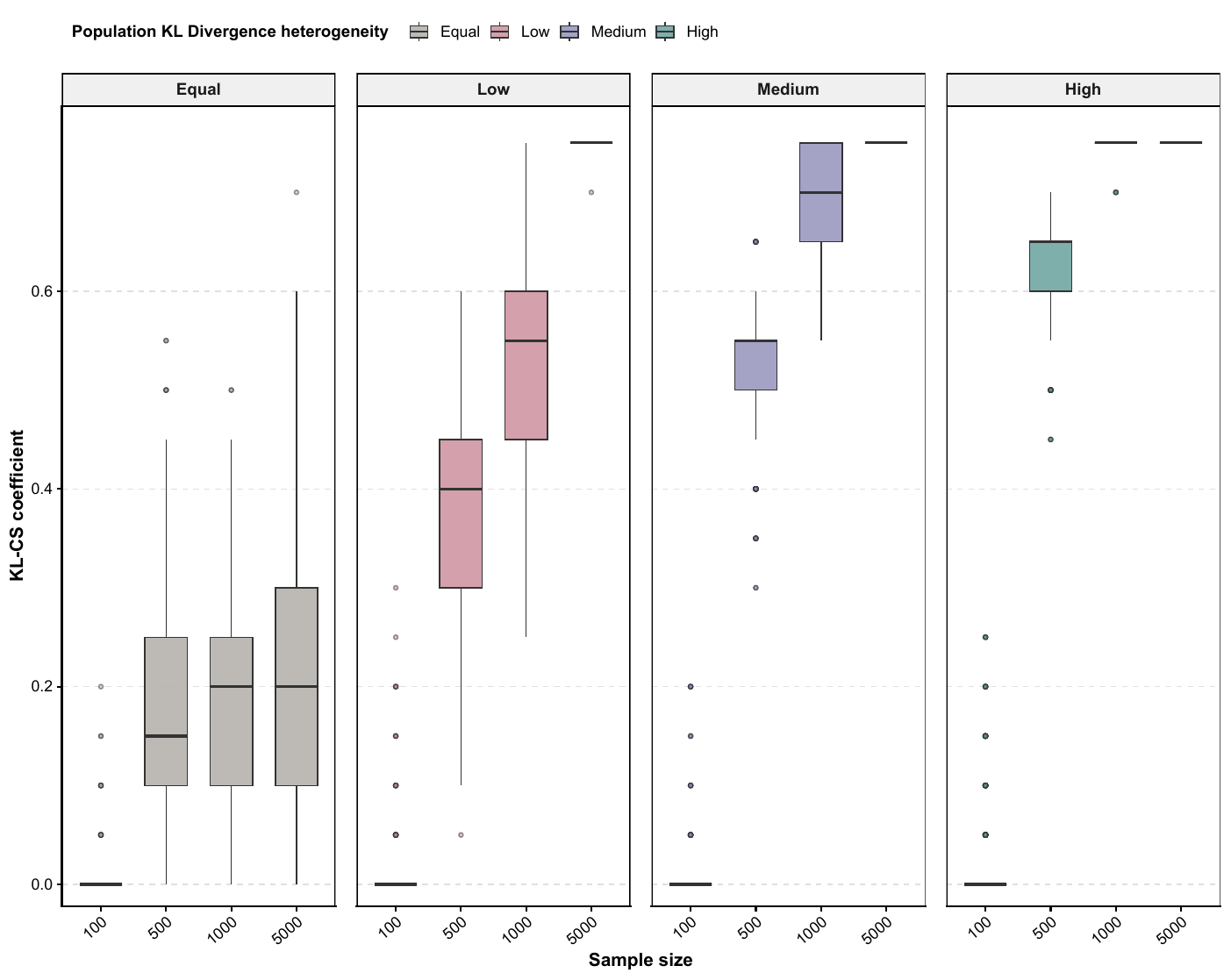}
\end{figure}

Figure~\ref{fig:kl_cs_simulation} presents the simulation results. With respect to the empirical reference thresholds, approximately 75\% of the CS coefficients under the Equal condition were below 0.25, whereas approximately 95\% were below 0.40. Accordingly, we propose 0.25 as the minimum reference threshold and 0.40 as the preferred threshold for the CS coefficient of KL divergence. Thus, substantive interpretation of between-node differences in KL divergence is not recommended when the CS coefficient is below 0.25, whereas a value above 0.40 is preferred. Regarding the performance of the CS coefficient, its values were close to zero across conditions when $n=100$. From $n=500$ onward, the CS coefficient generally increased with both sample size and population KL divergence heterogeneity, with mean values reaching 0.75 under the larger-sample conditions.

\subsection{Simulation Study on Bootstrap Difference Test}
Following the evaluation of GGMNIRA recovery and the establishment of preliminary reference thresholds for the CS coefficient of KL divergence, we next assessed the statistical validity of applying the bootstrap difference test to KL divergence. Specifically, this simulation study examined whether the test could adequately control the Type I error rate and how its statistical power changed across sample sizes and levels of population KL divergence heterogeneity.

The Type I error rate was defined as the proportion of node pairs with equal population KL divergence values that were incorrectly identified as significantly different and was therefore evaluated under the Equal condition. Statistical power was defined as the proportion of node pairs with a nonzero population difference in KL divergence that were correctly identified as significantly different and was therefore evaluated for the corresponding node pairs under the Low, Medium, and High conditions.

As in the CS-coefficient simulation, resampling was conducted within each simulated data generated under the common simulation framework. Here, however, 1,000 nonparametric bootstrap samples were obtained by resampling cases from the corresponding simulated data with replacement. For each bootstrap sample, the network was re-estimated using EBICglasso, and the corresponding KL divergence values were recalculated using GGMNIRA. For each of the 28 node pairs, the bootstrap distribution of the difference in KL divergence was then obtained, and a two-sided 95\% percentile confidence interval was constructed. When the confidence interval did not include zero, the equality of the KL divergence values for the node pair was rejected. 

The results indicated that the bootstrap difference test was approximately valid for KL divergence. The Type I error rates were 0, 0.0456, 0.0475, and 0.0522 for sample sizes of 100, 500, 1,000, and 5,000, respectively. Thus, the test was highly conservative at $n=100$, whereas the Type I error rate was close to the nominal significance level of $\alpha=.05$ when the sample size was 500 or larger. Statistical power generally increased with both sample size and population KL divergence heterogeneity. However, relatively high power still required a large sample size; for example, power exceeded 0.80 only at $n=5{,}000$ under the condition with the greatest heterogeneity. Overall, these results suggest that the bootstrap difference test provides approximately valid inference for KL divergence, although its power may be limited at smaller sample sizes.

\begin{figure}[!htbp]
    \centering
    \caption{Rejection rates of the bootstrap difference test for KL divergence across sample sizes and levels of population KL divergence heterogeneity. The rejection rate represents the Type I error rate under the Equal condition and statistical power for node pairs with nonzero population KL divergence differences under the Low, Medium, and High conditions. The horizontal dashed line indicates the nominal significance level of $\alpha=.05$.}
    \label{fig:kl_boot_simulation}
    \apafiguregraphic{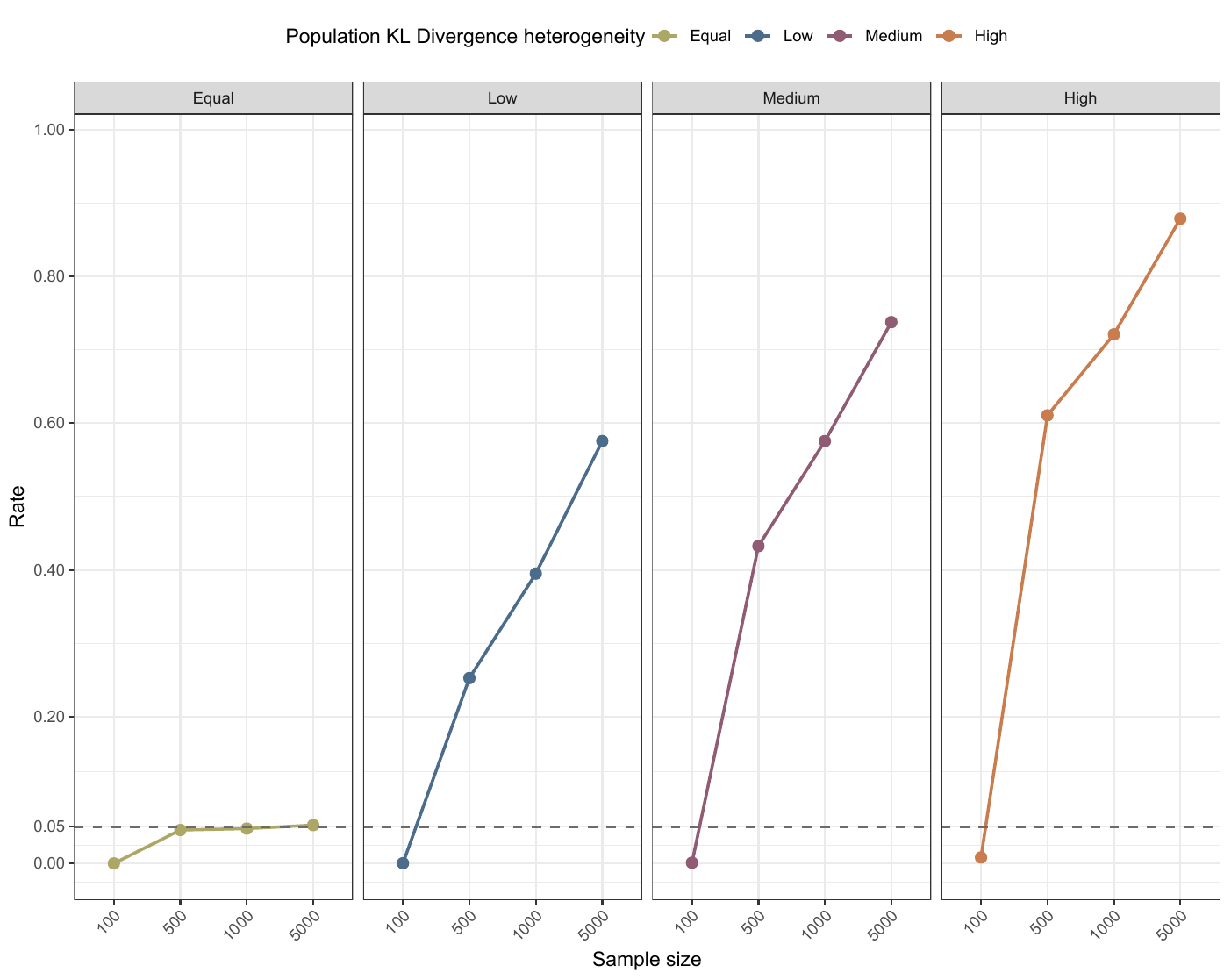}
\end{figure}

\section{Empirical Demonstration}
After introducing the complete GGMNIRA framework, we now provide an empirical data example and a reproducible R tutorial to illustrate its application in psychological network analysis. Specifically, using a Rest Intolerance dataset, we illustrate how GGMNIRA can be used to preliminarily identify which specific manifestation of rest intolerance is projected to be most important and may therefore warrant priority attention in future interventions. The R Markdown file and the rendered HTML report for reproducing the analysis are available at \url{https://github.com/ywu6888/GGMNIRA-reproducible-materials}.

\subsection{Loading the Required R Packages and Example Data}
GGMNIRA is implemented through our developed ``\Rpkg{GGMNIRA}'' R package. Before conducting the analysis, the required R packages need to be installed and loaded:

\begin{lstlisting}[style=RStyle]
# Install and load the corresponding packages
> install.packages("remotes")
> remotes::install_github("ywu6888/GGMNIRA")
> library(GGMNIRA)
\end{lstlisting}

\subsection{Introduction and Import of the Example Data}
The example data were drawn from a cross-sectional Rest Intolerance dataset collected by the authors, comprising 745 valid observations. Details of data collection and ethical approval are provided in the online supplementary materials. Participants' levels of Rest Intolerance were assessed using the Rest Intolerance Scale developed by \textcite{wang2025development}. The scale consists of eight items covering four dimensions: negative feelings (NF), social comparison (SC), cognitive bias (CB), and obsessive thinking (OT), with two items for each dimension. Items are rated on a 5-point Likert scale ranging from 1 (``strongly disagree'') to 5 (``strongly agree''), with higher total scores indicating higher levels of Rest Intolerance. The validity of this scale has been fully supported in multiple studies, like \textcite{wu2026daily}.

In addition, to clarify for readers whether the analysis results differ across data types under the NIRA framework, we recoded the example data to construct binary categorical variables; that is, original scores from 1 to 4 were coded as 0, whereas a score of 5 was coded as 1. We then fitted an Ising model and the corresponding NIRA to this binary dataset and compared the results with the GGMNIRA results.

We loaded these data in RStudio and named the dataset \Robj{RIdata}. The first six rows of the data are shown below, and the correspondence between the variable names and the specific items is provided in the Appendix~\ref{app:CT}.

\begin{lstlisting}[style=RStyle]
# View the first six rows of the Rest Intolerance data
> head(RIdata)
\end{lstlisting}
\begin{lstlisting}[style=ROutStyle]
# A tibble: 6 x 8
    NF1   SC1   OT1   CB1   NF2   SC2   OT2   CB2
1     3     3     3     3     3     3     3     3
2     4     4     3     4     3     2     2     3
3     2     3     4     5     4     4     5     5
4     1     2     2     2     2     2     2     2
5     3     3     3     4     2     3     3     3
6     3     3     4     4     3     3     4     4
\end{lstlisting}

\subsection{Network Estimation and Implementation of GGMNIRA}
\begin{figure}[!htbp]
    \centering
    \caption{Gaussian graphical model of Rest Intolerance. The network was estimated from a cross-sectional dataset with 745 valid observations. Nodes represent the eight items of the Rest Intolerance Scale, covering negative feelings (NF), social comparison (SC), cognitive bias (CB), and obsessive thinking (OT). }
    \label{fig:rigraph}
    \apafiguregraphic{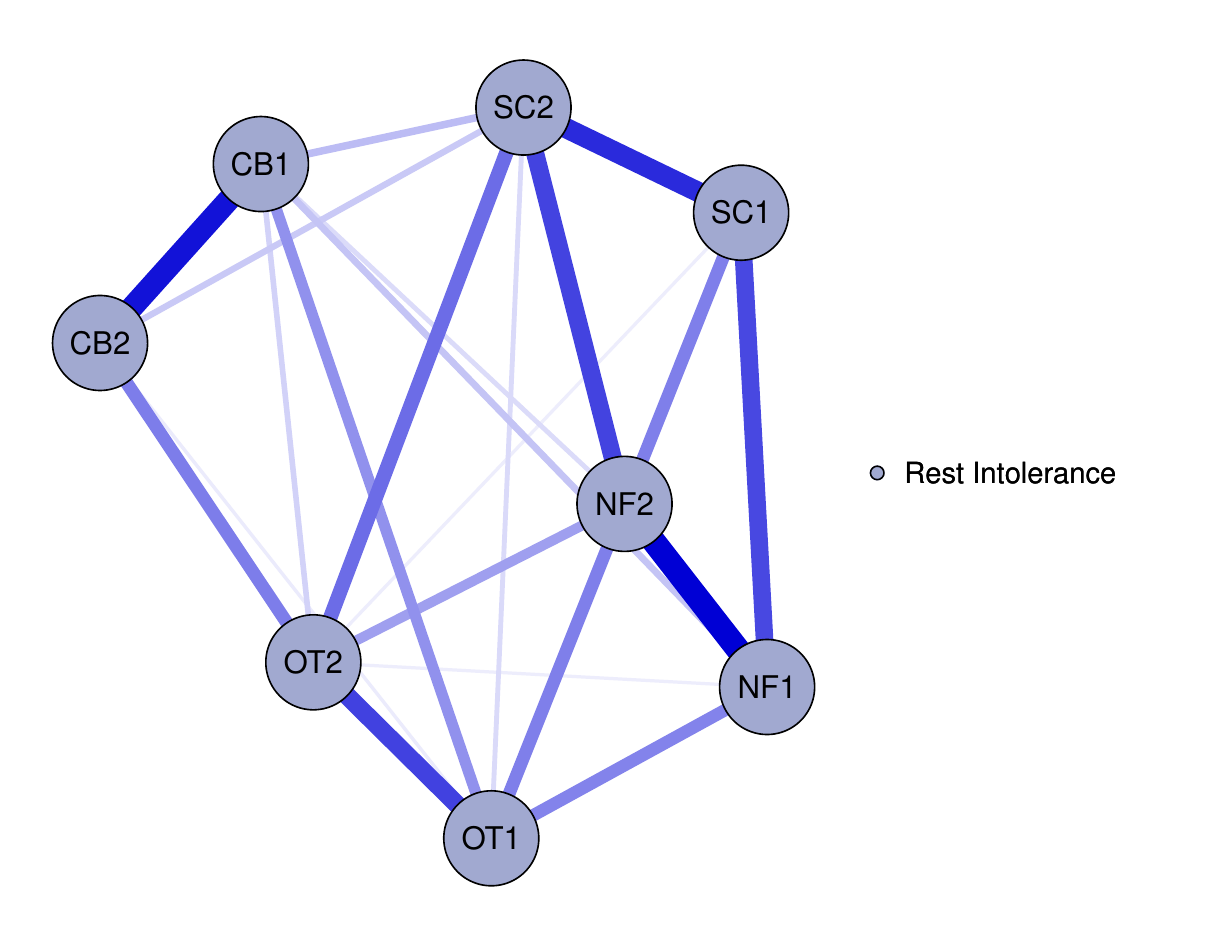}
\end{figure}

To characterize the conditional dependence relationships among the 8 Rest Intolerance items, we estimated a regularized Gaussian graphical model using the \Rfunc{estimateNetwork} function. The model selection algorithm was set to ``\Rarg{EBICglasso}'', and the correlation input method was set to ``\Rarg{cor_auto}'', thereby using a polychoric correlation matrix to estimate the network structure. The resulting network is shown in Figure~\ref{fig:rigraph}.

\begin{lstlisting}[style=RStyle]
# Estimated Regularized Gaussian Graph Model
> net <- bootnet::estimateNetwork(RIdata,default = "EBICglasso",corMethod = "cor_auto")
# Plot a visual network graph
> Plot_network <- plot(net, layout = 'spring', 
                                groups=rep(c("Rest Intolerance"),8),
                                color=c("#A1A9D0"),legend=TRUE)
\end{lstlisting}

After obtaining the regularized Gaussian graphical model for Rest Intolerance, we applied \Rfunc{GGM_NIRA} to compare the projected importance of the eight nodes and identify which node was projected to produce the greatest network-level change when manipulated. Specifically, we input the network object \Robj{net} into the \Rfunc{GGM_NIRA} function and used the \Rarg{type} argument to specify that all nodes in this network belong to the Rest Intolerance construct. The \Rfunc{GGM_NIRA} function automatically computes the minimum manipulation coefficient for practical application that is required for the manipulation to be effective\footnote{Because the manipulation direction does not change the results, the manipulation direction is set to be positive by default.}, and then performs simulated manipulation under this manipulation coefficient. Users may also manually specify the manipulation coefficient, but in practical applications we recommend prioritizing the automatic mode. Subsequently, the \Rfunc{GGMNIRA_result} function can be used to output the simulated manipulation results for each node.

\begin{lstlisting}[style=RStyle]
# Input the regularized GGM for GGMNIRA analysis
> RI_res <- GGM_NIRA(net = net, type = rep(c("Rest Intolerance"),8))
# Output GGMNIRA results
> GGMNIRA_result(RI_res)
\end{lstlisting}

After running the above code, the \Rfunc{GGM_NIRA} function outputs the spectral radius of the regression coefficient matrix $B$, the threshold for KL divergence, the minimum manipulation coefficient for practical application, and the first node that reaches the threshold. The \Rfunc{GGMNIRA_result} function further summarizes the names of the manipulated nodes and their corresponding minimum manipulation coefficients and KL divergences.

Results show that NF2, reflecting feeling like a loser during rest, had the largest KL divergence. This suggests that NF2 may represent the manifestation of Rest Intolerance with the greatest projected importance and may therefore warrant priority attention as a potential target for future intervention.

\begin{lstlisting}[
    style=ROutStyle,
    breaklines=true,
    breakatwhitespace=true,
    basicstyle=\ttfamily\small,
    columns=fullflexible
]
# Spectral radius rho(B) = 0.948467 (rho != 1)
# Node-level joint KL: joint dimension p = 8, alpha = 0.0500, joint KL threshold = 7.753657
# Automatic integer manipulation intensity for Node-level joint KL = 1.000000 (trigger node: NF2)
# A data frame: 8 x 3
  manipulated_node required_manipulation_I       KL
1              NF2               0.9916344 7.885032
2              SC2               1.0074149 7.639937
3              NF1               1.0746957 6.713292
4              SC1               1.1116750 6.274093
5              OT1               1.1819031 5.550637
6              OT2               1.2019094 5.367390
7              CB1               1.6057219 3.007225
8              CB2               1.9454818 2.048577
\end{lstlisting}

To more intuitively illustrate the simulated manipulation effects of each node, we can visualize the results using the \Rfunc{GGMNIRA_plot} function, in which the \Rarg{manipulation_I} argument was used to specify the minimum manipulation coefficient for practical application. Based on the output messages provided by the \Rfunc{GGM_NIRA} function, the current minimum manipulation coefficient for practical application is 1; therefore, we set \Rarg{manipulation_I} = 1 here.

\begin{lstlisting}[style=RStyle]
# Visualization of GGMNIRA Result
> GGMNIRA_Plot <- GGMNIRA_plot(RI_res, manipulation_I = 1)
> print(GGMNIRA_Plot)
\end{lstlisting}

\begin{figure}[!htbp]
    \centering
    \caption{Projected effect plot after manipulation using the GGMNIRA. Note: The x-axis represents the manipulated nodes, and the y-axis represents the KL divergence between the manipulated distribution and the baseline distribution. An asterisk (*) indicates that, at the current manipulation intensity, manipulating the corresponding node produces an effective effect.}
    \label{fig:rikl}
    \apafiguregraphic{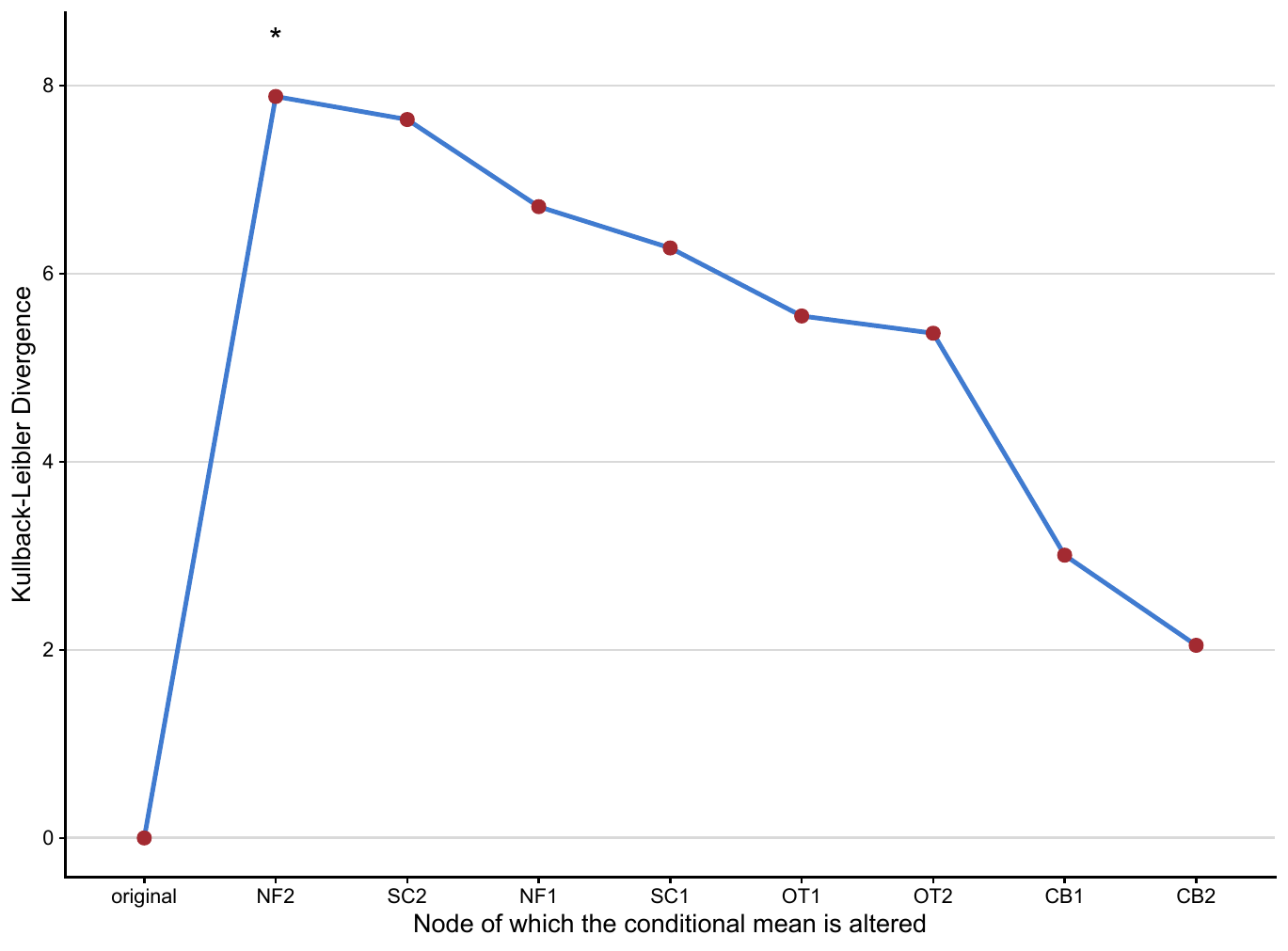}
\end{figure}

After completing the computation of the main function described above, the stability of the GGMNIRA results should be further evaluated. For this purpose, we provide the \Rfunc{bootnet_KL} function to perform case-dropping bootstrap analyses. Users only need to input the data, information on the network construct, the number of bootstrap replications, and the random seed. Because changes in the manipulation intensity and direction do not affect the stability estimation results, the \Rfunc{bootnet_KL} function sets the manipulation coefficient to 1 by default, and users do not need to specify it manually.

\begin{lstlisting}[style=RStyle]
> boot_result <- bootnet_KL(data = RIdata, type = rep(c("Rest Intolerance"),8), nBoots = 1000, nCores = 8, seed = 2026)
\end{lstlisting}

We can then compute the CS coefficient for KL divergence using the \Rfunc{corStability_KL} function and plot the stability results using the \Rfunc{plot_KL} function.

\begin{lstlisting}[style=RStyle]
> CS_result <- corStability_KL(boot_result = boot_result)
> plot_KL(CS_result = CS_result, boot_result = boot_result)
\end{lstlisting}

\begin{figure}[!htbp]
    \centering
        \caption{Average correlations between KL divergence of networks sampled with persons dropped and the original sample. Lines indicate the means and areas indicate the range from the 2.5th quantile to the 97.5th quantile.}
        \label{fig:riklcs}
    \apafiguregraphic{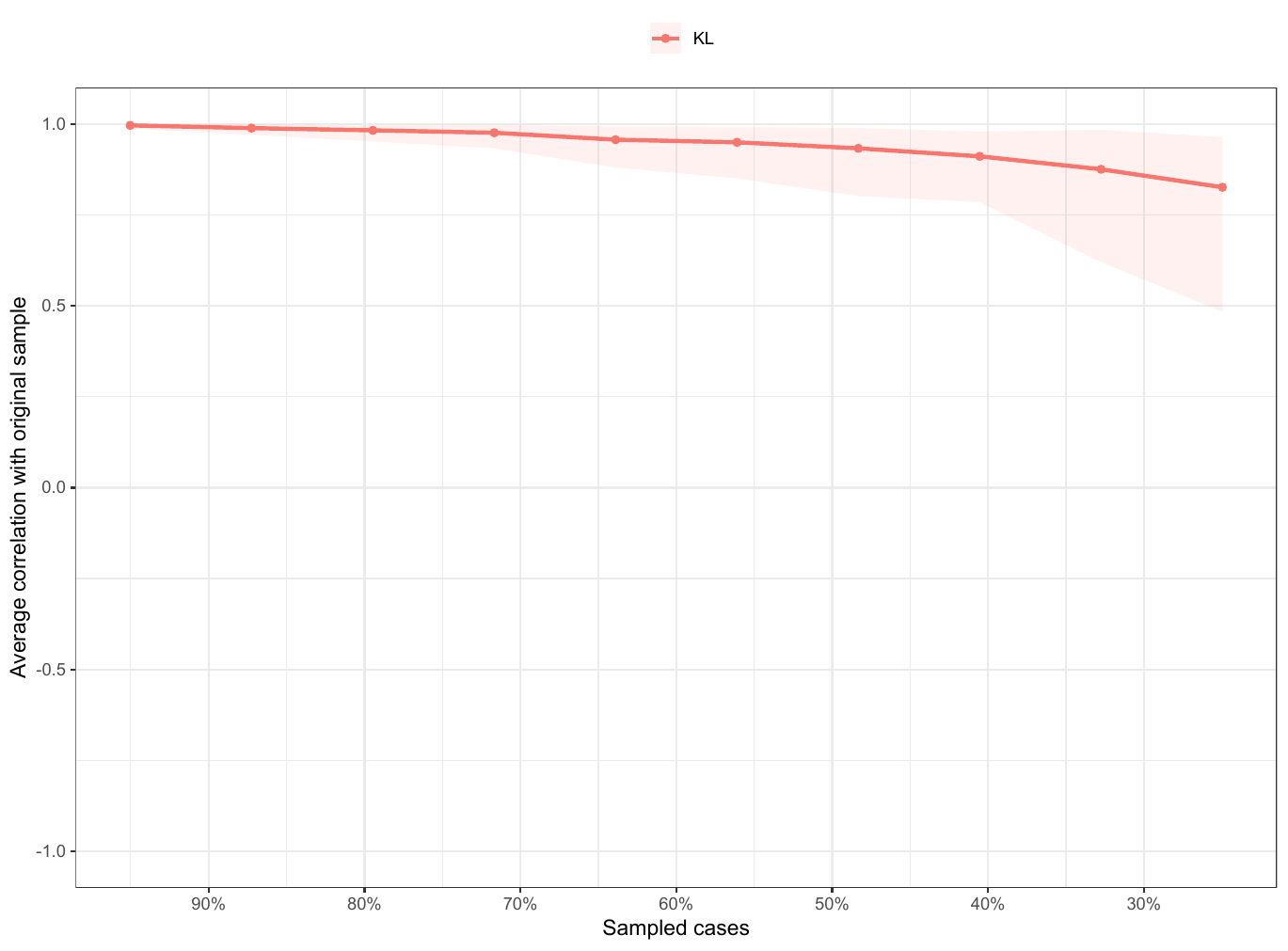}
\end{figure}

The results show that the CS coefficient for KL divergence is 0.672, which is higher than the preliminary threshold of 0.40 identified in our simulation study. This indicates that the KL divergence in the present study has a certain degree of stability.

Furthermore, we can also conduct significance tests for differences in KL divergence between nodes.  Specifically, nonparametric bootstrap sampling can be performed via the \Rfunc{bootnet_KL_nonparametric} function and, in conjunction with the \Rfunc{differenceTest_KL} function, 95\% confidence intervals for pairwise differences in KL divergence between nodes can be constructed, thereby assessing whether the KL divergences of different nodes are statistically distinguishable. Correspondingly, the \Rfunc{plot_KL_difference} function can be used to visualize the bootstrap difference test results, providing an intuitive display of pairwise differences in KL divergence between nodes and their significance patterns.

It should be noted that, unlike the \Rfunc{bootnet_KL} function, the \Rfunc{bootnet_KL_nonparametric} function still requires users to manually specify the manipulation coefficient \Rarg{manipulation_I}. This is because, although changes in manipulation intensity likewise do not affect the statistical conclusions regarding pairwise KL divergence differences in the bootstrap difference test, the resulting difference test plot displays the actual KL divergence estimates. Therefore, when conducting the nonparametric bootstrap difference test, users should specify the corresponding \Rarg{manipulation_I} based on the manipulation coefficient message output by the \Rfunc{GGM_NIRA} function, so as to ensure that the KL divergence displayed in the figure are consistent with the results of the main analysis.

\begin{lstlisting}[style=RStyle]
> boot_nonparametric_result <- bootnet_KL_nonparametric(
   data = RIdata,
   net = net,
   type = rep(c("Rest Intolerance"),8),
   nBoots = 1000,
   manipulation_I = 1,
   seed = 2026)
> diff_result <- differenceTest_KL(boot_nonparametric_result)
> plot_KL_difference(
   diff_result = diff_result,
   sample_KL   = boot_nonparametric_result$sample_KL,
   order       = "sample"
 )
\end{lstlisting}

\begin{figure}[!htbp]
    \centering
        \caption{Bootstrap difference tests ($\alpha$ = 0.05) were conducted among the KL divergence of the eight rest Intolerance manifestations in the estimated network. Grey boxes indicate nodes with insignificant differences in KL divergence from each other, black boxes represent nodes with significant differences in KL divergence from each other, and white boxes show the values of KL divergence of the nodes.}
        \label{fig:riklboot}
    \apafiguregraphic{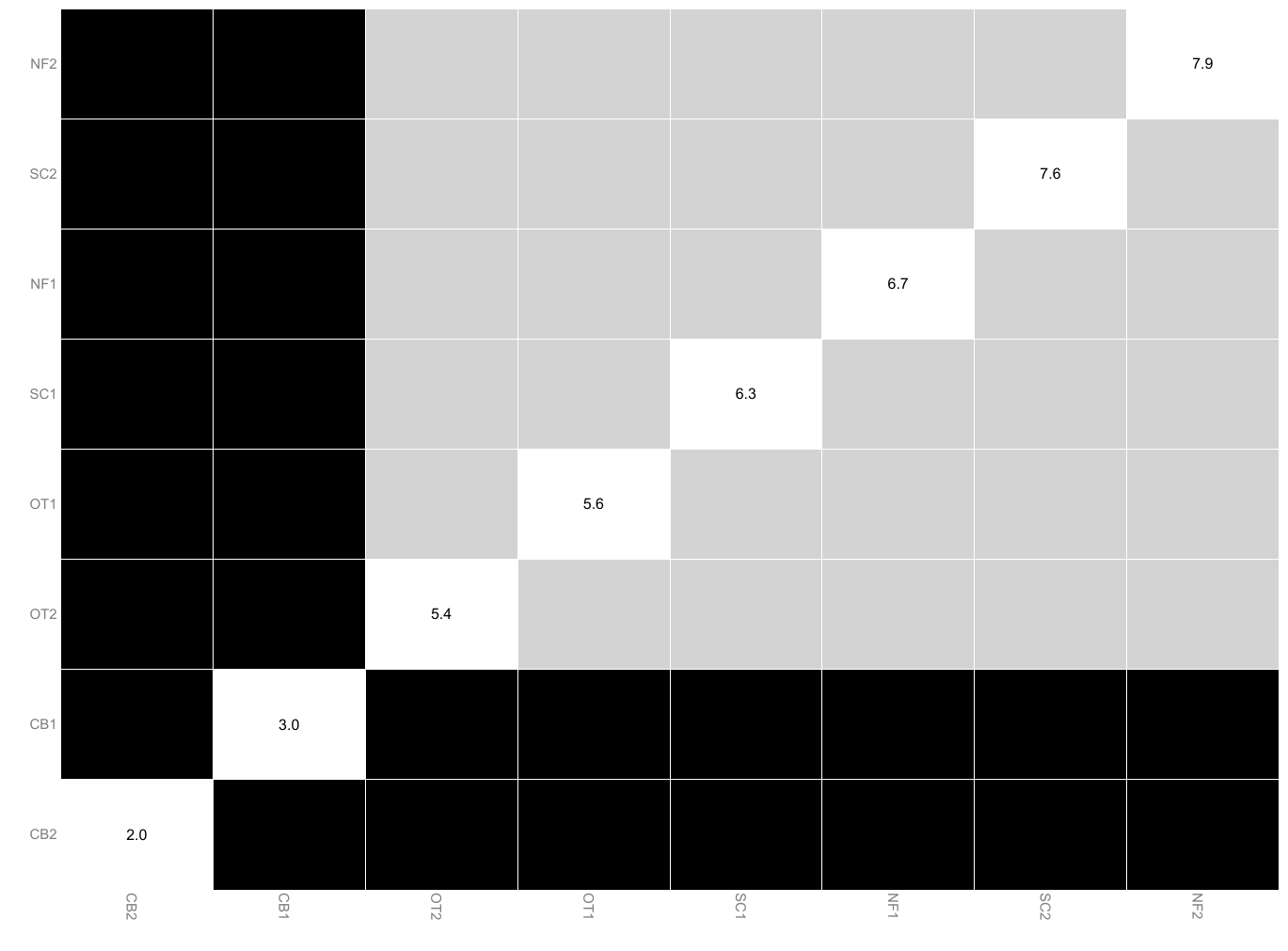}
\end{figure}

The results show that the KL divergence of CB1 and CB2 differed significantly from those of all other manifestations, and that the KL divergence of CB1 and CB2 also differed significantly from each other. It is important to note that the sample size of the present study is not sufficient to support the statement that ``the KL divergence of all manifestations except CB1 and CB2 are equal.'' Rather, we can only state that ``the KL divergence of all manifestations except CB1 and CB2 did not differ significantly.''

\subsection{Compared with the original NIRA results based on binarized data}
As discussed above, to compare the results of original NIRA and GGMNIRA, we binarized the data, fitted an Ising model using the \Rfunc{estimateNetwork} function, and visualized the network.

\begin{lstlisting}[style=RStyle]
# Define a function for binary conversion
> to_01_from_5pt <- function(x, cutpoint = 5) {
  x_num <- as.integer(as.character(x))  
  as.integer(x_num >= cutpoint)
}
> RIdata_binary <- as.data.frame(lapply(RIdata, to_01_from_5pt, cutpoint = 5))
# Estimated Regularized Ising Model
> RI_ising <- bootnet::estimateNetwork(RIdata_binary,default = "IsingFit")
# Plot a visual network graph
> graph <- plot(RI_ising, layout = 'spring',
                            groups=rep(c("Rest Intolerance"),8),
                            color=c("#F9EB77"),legend=TRUE)
\end{lstlisting}

We then conducted the original NIRA analysis following the tutorial by \textcite{wang2026simulation}. We do not provide an extensive discussion of the code details here; readers may refer to the Online R Markdown document for further details. 

\begin{figure}[!htbp]
    \centering
    \caption{Ising network and original NIRA results for Rest Intolerance. Panel A presents the Ising network estimated from dichotomized item responses. Panels B and C show the original NIRA results under alleviating and aggravating interventions, respectively.}
    \label{fig:riisingnira}
    \apafiguregraphic{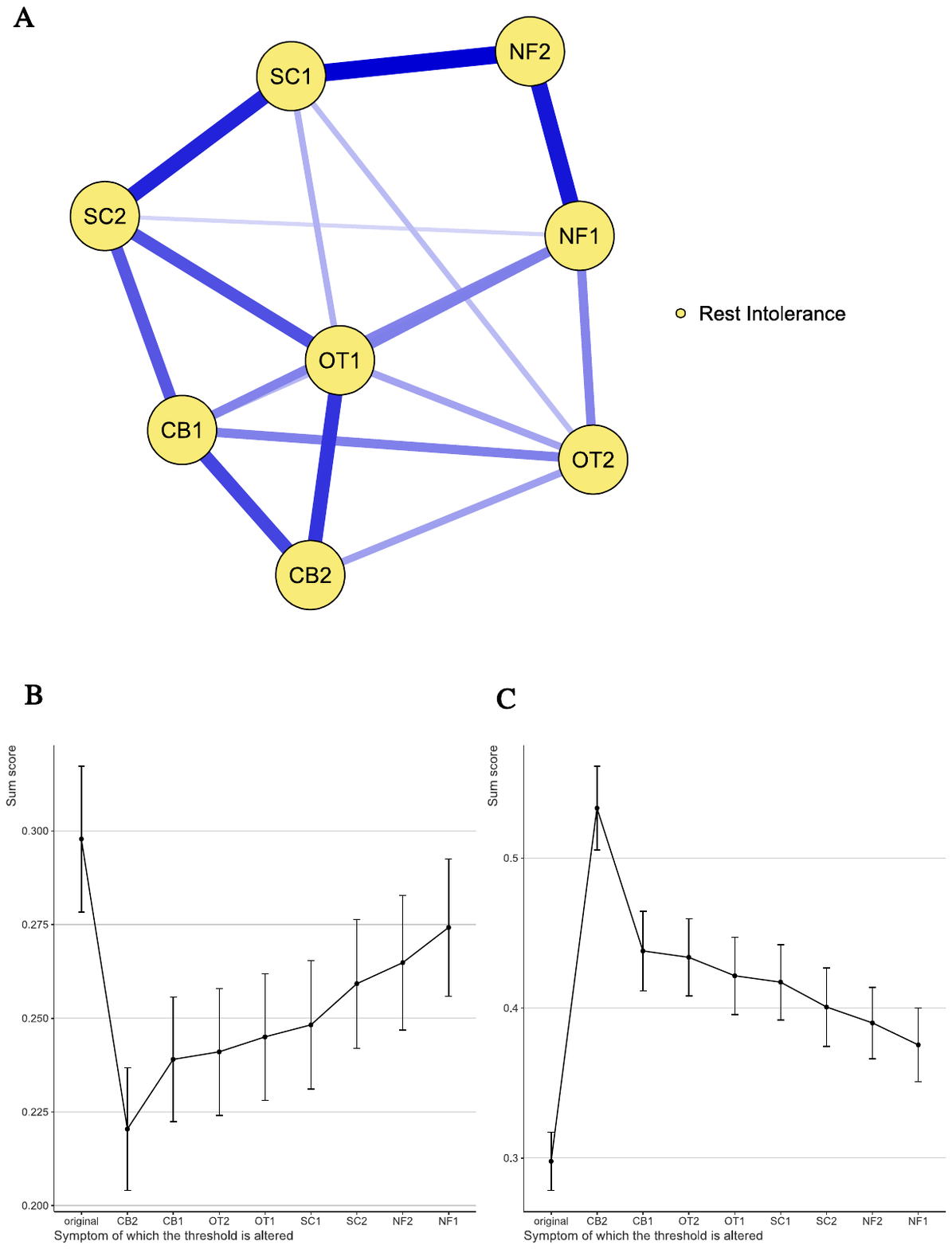}
\end{figure}

The results showed that, for both alleviating and aggravating interventions, CB2 was the node with the greatest projected importance in the network. This differs from the node identified as having the greatest projected importance by GGMNIRA, indicating that conclusions obtained by forcibly dichotomizing ordinal data for analysis may not necessarily be reliable.

\section{Extending GGMNIRA to Bridge Network}
In the preceding sections, we introduced the GGMNIRA framework and illustrated its application in psychological network analysis. However, these developments essentially focus on networks involving a single psychological construct. Many research questions in psychology, by contrast, are not limited to the relationships among variables within a single construct, but instead concern the interconnections among multiple constructs, such as those between depression and anxiety or between stress and sleep. When variables representing different theoretical constructs are included in the same network and estimated jointly, the resulting network is commonly referred to as a \textit{bridge network} \parencite{jones2021bridge}.\footnote{In this article, we sometimes refer to this complete multi-construct network as a ``joint network'' to emphasize that it contains all variables from all constructs. When the bridge network does not specifically refer to local connection structures between particular constructs, the two terms can be used interchangeably.}

Bridge networks have been widely applied across different areas of psychology. For example, in psychopathology, symptoms from different disorders are often included in the same network to examine comorbidity, with variables connecting different symptom communities conceptualized as bridge symptoms \parencite{cramer2010comorbidity, borsboom2013network}. In educational psychology, variables from multiple learning-related constructs have been jointly modeled to identify key variables involved in learning adaptation \parencite{putwain2024revisiting}. Similarly, in personality and vocational psychology, cross-construct networks have been used to examine how personality traits and career-related factors are interconnected \parencite{wang2024network}. Given the broad application and substantive importance of bridge networks, systematically extending GGMNIRA from single-construct networks to bridge networks has clear methodological significance and practical value. Accordingly, in this section, we introduce the Bridge Gaussian Graphical Model NodeIdentifyR Algorithm (BGGMNIRA) and describe how GGMNIRA can be adapted to address research questions involving multiple constructs.

We propose that GGMNIRA can be extended to bridge networks at three levels. These levels correspond to three types of analytical questions with distinct research aims.

The first level is node-level GGMNIRA from the joint-network perspective. Conceptually, this level is identical to the single construct case described in the previous section. Specifically, the algorithm sequentially manipulates each node across all constructs in the joint network. It then computes the KL divergence between the joint distribution of the complete manipulated network and the baseline distribution. This level therefore answers the following question: for the entire bridge network, which node is projected to produce the greatest overall change after being manipulated? In a depression-anxiety comorbidity network, for example, this level can be used to answer which symptom has the greatest projected importance in the overall comorbidity network. This level of analysis is suitable when researchers aim to obtain a comprehensive ranking of node projected importance across all constructs.

The second level is node-level GGMNIRA from the construct-specific perspective.\footnote{The simplified expression for KL divergence in Equation~\eqref{eq:11} was derived for node-level manipulation when the distributional change is evaluated at the level of the complete joint network. It therefore does not directly apply to the second and third levels of bridge GGMNIRA. The corresponding mathematical derivations and detailed explanations are provided in the online supplementary materials.} In this analytical framework, researchers need to explicitly specify a source construct and a target construct. The algorithm sequentially manipulates each node in the source construct. However, it no longer evaluates the overall distributional change at the level of the joint network. Instead, it examines only the extent to which the marginal distribution of the target construct changes after manipulation. Thus, this level answers the following question: which node in the source construct is projected to have the largest impact on the target construct after being manipulated? Returning to the depression-anxiety comorbidity network as an example, if depression is specified as the source construct and anxiety as the target construct, this level can be used to answer which depressive symptom is projected to produce the largest change in anxiety after manipulation. This can help identify key comorbid symptoms with cross-disorder influence.

The third level is construct-level GGMNIRA from the joint-network perspective. In contrast to the node-level manipulations in the first two levels, the target of manipulation at this level is the construct as a whole. That is, the algorithm applies manipulations of equal intensity and direction simultaneously to all nodes belonging to the same construct, thereby implementing simulated manipulation at the construct level. It then sequentially iterates over each construct and computes the KL divergence between the joint distribution of the complete manipulated network and the baseline distribution. It is important to note that the number of nodes may differ across constructs. If the same manipulation intensity is applied to every node, the effective manipulation intensity at the construct level will not be equivalent across constructs. For example, if construct $A$ contains 9 nodes and construct $B$ contains 7 nodes, and a unit-intensity manipulation is applied to all 16 nodes, then the effective manipulation intensity at the construct level is 9 for $A$ and 7 for $B$, resulting in a systematic difference between the two constructs. To ensure comparability across constructs, we impose a unit-norm constraint to normalize the manipulation vector, thereby ensuring consistency in manipulation intensity at the construct level.

It is worth noting that construct-level GGMNIRA has clear practical significance. Taking psychopathology as an example, in actual psychological interventions, researchers and clinicians may question whether intervention hypotheses at the level of symptoms are overly idealized. This is because interventions in real world often do not change only one isolated symptom within a disease or disorder, but instead intervene directly on the disease or disorder itself. This has also been referred to by researchers as a ``fat-handed intervention'' \parencite{bringmann2019centrality, eronen2020causal}. Construct-level manipulation is proposed in response to this need, as it makes the hierarchical structure of simulated manipulation more consistent with the logic of clinical practice. However, it should again be emphasized that all of the manipulations described above are statistical simulations based on model parameters. Their results cannot be directly equated with the effects of real clinical interventions, and should only be used as a reference for practical decision making.

For uncertainty assessment, we also extend the stability estimation and difference testing framework used for single construct GGMNIRA to BGGMNIRA. Specifically, we continue to use the CS coefficient to evaluate the stability of BGGMNIRA results. We also use the bootstrap difference test to examine significant differences in KL divergence between nodes or between constructs in the bridge network, thereby supporting valid inference about the estimated results.

In addition, we also provide a example dataset about depression and anxiety to demonstrate the specific procedure for implementing BGGMNIRA. For reasons of length and readability, the R Markdown file and the rendered HTML report for reproducing the analysis are available at
\url{https://github.com/ywu6888/BGGMNIRA-reproducible-materials}, and readers may consult it directly.

\section{Discussion}
\label{sec:dis}
The value of NIRA lies not only in providing a manipulation-based alternative for evaluating node importance in psychological networks, but also in its potential practical utility. For psychological variables that are difficult, impossible, or ethically inappropriate to manipulate directly in the real world, such as childhood adversity, researchers can use NIRA to examine the projected outcomes of simulated manipulations. Although these outcomes remain model-generated expectations and cannot substitute for real-world causal evidence, they can still provide a basis for formulating theoretical hypotheses to be examined in subsequent empirical research. Despite these advantages, NIRA remains limited to binary variables, and the theoretical interpretation of the node intercept it manipulates is primarily grounded in the psychopathology.

Building on NIRA, the present study extends its logic of simulated manipulation to Gaussian graphical model and proposes GGMNIRA. Fundamentally, the two approaches can be viewed as specific implementations of a broader simulated manipulation framework, as both implement simulated manipulations by perturbing the intercept in a node’s conditional model. The key difference lies in their interpretive emphasis. Whereas the original NIRA focuses primarily on changes in the intercept, GGMNIRA places greater emphasis on the corresponding change in the node’s conditional mean, thereby allowing simulated manipulation to be applied to a broader range of psychological variables and research contexts. Specifically, GGMNIRA systematically changes the conditional mean of a given node and uses KL divergence to quantify the degree of difference between the manipulated network distribution and the baseline distribution, thereby evaluating the projected importance of each node. Although Equation~\eqref{eq:16} shows that, in the basic single-network setting, GGMNIRA results have a mathematical correspondence with node predictability, it is important to emphasize that GGMNIRA is better viewed as a foundational framework for simulated manipulation. Its broader value lies in providing a basis for the development of additional analytical procedures and methodological extensions. For example, in the present study, we developed complementary procedures for evaluating the uncertainty of GGMNIRA estimates.

We further extended GGMNIRA to bridge network, demonstrating its broader applicability as a framework for simulated manipulation. This extension supports simulated manipulation at both the node and construct levels, allowing researchers to select the appropriate level of analysis according to their specific research questions. In particular, construct-level manipulation may have greater practical relevance in some real-world settings, because actual manipulations may affect a group of related nodes rather than a single isolated node. Questions concerning changes in a specific target construct and construct-level manipulation extend beyond what can be addressed by node predictability alone.

The applicability of GGMNIRA to ordinal data also requires clarification. Because the underlying graphical model of GGMNIRA is the GGM, strictly speaking, the algorithm is directly applicable to continuous Gaussian variables. However, for ordinal variables such as Likert-type responses, psychological network research commonly adopts a latent Gaussian threshold model, in which the observed ordinal categories are assumed to arise from underlying continuous Gaussian variables through a thresholding process, and polychoric correlations are used to estimate the corresponding latent correlation structure \parencite{epskamp2018tutorial}. When this latent Gaussian assumption is reasonable, GGMNIRA can also be applied to such ordinal data. Nevertheless, this does not imply that the current method is generally applicable to all forms of ordinal data. Developing NIRA methods for more general ordinal data therefore remains an important direction for future research.

Several limitations of GGMNIRA should also be considered. First, GGMNIRA evaluates projected node importance rather than the actual causal effect of a real-world manipulation. Its results should therefore be understood as theoretical results generated by the model and requiring empirical validation, rather than as established causal conclusions \parencite{guest2021computational,robinaugh2021invisible}. Second, the current GGMNIRA is developed within a linear framework. In real psychological systems, however, relationships among variables may be more complex. For example, when moderation effects are present, manipulating one variable may also alter the relationships among other variables in the network. The current GGMNIRA cannot fully accommodate such situations. A natural extension is therefore to develop the simulated manipulation logic of GGMNIRA within the moderated Gaussian graphical model framework, so that changes in network relationships accompanying node manipulation can also be incorporated. We are currently pursuing this direction. Finally, the current GGMNIRA is primarily developed for cross-sectional, group-level networks, and its results should therefore not be directly interpreted as within-person dynamic processes over time. Future research may extend the framework to temporal or personalized network models based on intensive longitudinal data. 

\section{Conclusion}
The present study extends the simulated-manipulation logic of NIRA to Gaussian graphical model and develops the GGMNIRA for evaluating projected node importance in psychological networks. Building on this core algorithm, we further incorporated procedures for uncertainty assessment and extended GGMNIRA to bridge Gaussian graphical model, thereby broadening the framework to multi-construct networks. Overall, GGMNIRA should not be viewed merely as a method for assessing node importance, but rather as a more general algorithmic framework. We hope that this framework will provide a useful foundation for further methodological development and for broader applications of simulated manipulation in psychological network research.

\printbibliography

\appendix

\section{Correspondence Table of Nodes and Items in the Example Data}
\label{app:CT}

\begin{table}[htbp]
\centering
\caption{Variable Names and Corresponding Items}
\label{tab:variables_items}

\begin{threeparttable}
\footnotesize
\setlength{\tabcolsep}{4pt}

\begin{tabularx}{\linewidth}{@{}l>{\raggedright\arraybackslash}X@{}}
\toprule
Variable & Item \\
\midrule

NF1 & ``When I am resting or having fun, I feel guilty.'' \\

NF2 & ``When I am resting or having fun, I feel like a loser.'' \\

OT1 & ``When I'm resting or having fun, I can't help but think about studying or working.'' \\

OT2 & ``When I'm resting or having fun, I always feel that there's something else I'm not doing.'' \\

CB1 & ``I think I should need to spend more time studying and working rather than resting.'' \\

CB2 & ``I thought I should use my time off to do something more meaningful.'' \\

SC1 & ``When I'm resting or having fun, I worry about being passed by someone else.'' \\

SC2 & ``When I take a break, I always think of my peers working harder than me.'' \\

\bottomrule
\end{tabularx}

\begin{tablenotes}[flushleft]
\small
\item \textit{Note.} NF = negative feelings; OT = obsessive thoughts; CB = cognitive bias; SC = social comparison.
\end{tablenotes}

\end{threeparttable}
\end{table}

\end{document}